\documentclass[11pt]{article}
\usepackage{graphicx}
\usepackage[margin=1.25in]{geometry}
\usepackage[usenames,dvipsnames]{color}
\usepackage{url}
\usepackage[colorlinks = true,
            linkcolor = blue,
            urlcolor  = blue,
            citecolor = blue,
            anchorcolor = blue]{hyperref}

\usepackage{cite}
\usepackage{wrapfig}
\usepackage{amsmath}
\usepackage{amssymb}
\usepackage{authblk}
\usepackage{enumitem}
\setlist[itemize]{noitemsep}

\makeatletter
\let\@fnsymbol\@alph
\makeatother

\newenvironment{Abstract}{\begin{quotation} \begin{center}
                       ABSTRACT
     \end{center}\bigskip  }{\end{quotation}}

\def\beq{\begin{equation}}
\def\eeq#1{\label{#1}\end{equation}}
\def\eeqn{\end{equation}}

\newenvironment{Eqnarray}%
   {\arraycolsep 0.14em\begin{eqnarray}}{\end{eqnarray}}
\def\beqa{\begin{Eqnarray}}
\def\eeqa#1{\label{#1}\end{Eqnarray}}
\def\eeqan{\end{Eqnarray}}

\let\bar=\overbar

\def\lsim{\mathrel{\raise.3ex\hbox{$<$\kern-.75em\lower1ex\hbox{$\sim$}}}}
\def\gsim{\mathrel{\raise.3ex\hbox{$>$\kern-.75em\lower1ex\hbox{$\sim$}}}}

\def\del{\partial}
\def\Dslash{\not{\hbox{\kern-4pt $D$}}}
\def\dslash{\not{\hbox{\kern-2pt $\del$}}}
\def\pslash{\not{\hbox{\kern-2pt $p$}}}
\def\ETmiss{\not{\hbox{\kern-4pt $E$}}_T}

\def\Dlr{\mathrel{\raise1.5ex\hbox{$\leftrightarrow$\kern-1em\lower1.5ex\hbox{$D$}}}}

\def\MSB{{\bar{M \kern -2pt S}}}
\def\msb{{\bar{\scriptsize M \kern -1pt S}}}

\def\drb{{\bar{\scriptsize D \kern -1pt R}}}

\newcommand\snowmass{\begin{center}\rule[-0.2in]{\hsize}{0.01in}\\\rule{\hsize}{0.01in}\\
\vskip 0.1in Submitted to the  Proceedings of the US Community Study\\ 
on the Future of Particle Physics (Snowmass 2021)\\ 
\rule{\hsize}{0.01in}\\\rule[+0.2in]{\hsize}{0.01in} \end{center}}

\begin{document}

\title{{\bf Snowmass White Paper:
            Precision Studies of Spacetime Symmetries
            and Gravitational Physics}\\
       \normalsize\snowmass}

\renewcommand\Affilfont{\itshape\normalsize}

\author[1]{Eric Adelberger}
\author[2]{Dmitry Budker}
\author[3]{Ron Folman}
\author[,4]{Andrew A. Geraci\thanks{email: andrew.geraci@northwestern.edu; corresponding author Sec.~\ref{ShortRangeBSM} and~\ref{GQI}}}
\author[,5]{Jason T.\ Harke\thanks{email: harke2@llnl.gov; corresponding author Sec.~\ref{Th}}}
\author[,6]{Daniel M.\ Kaplan\thanks{email: kaplan@iit.edu; corresponding author Sec.~\ref{section-mage}}}
\author[7]{Derek F. Jackson Kimball}
\author[,8]{Ralf Lehnert\thanks{email: ralehner@indiana.edu; corresponding author Sec.~\ref{CPT}}}
\author[9]{David Moore}
\author[10]{Gavin W.\ Morley}
\author[,11]{Anthony Palladino\thanks{email: palladin@bu.edu; corresponding author Sec.~\ref{section-kaon-system}}}
\author[12]{Thomas J.\ Phillips
}
\author[12,13,14]{Giovanni M.\ Piacentino
}
\author[,8,15]{William Michael Snow\thanks{email: wsnow@indiana.edu; corresponding author Sec.~\ref{NOPTREX} and \ref{exotic}}}
\author[,16]{Vivishek Sudhir\thanks{email: vivishek@mit.edu; corresponding author Sec.~\ref{GQI}}}

\affil[1]{Department of Physics, 
Center for Experimental Nuclear Physics and Astrophysics,
University of Washington
Seattle, WA 98195 USA}
\affil[2]{Helmholtz Institute, JGU Mainz and UC Berkeley}
\affil[3]{Ben-Gurion University of the Negev, Israel}
\affil[4]{Center for Fundamental Physics, Northwestern University, Evanston, IL 60208, USA}
\affil[5]{Lawrence Livermore National Laboratory, Livermore, CA, USA}
\affil[6]{Illinois Institute of Technology, Chicago, IL, USA}
\affil[7]{California State University - East Bay}
\affil[8]{Indiana University Center for Spacetime Symmetries, Bloomington, Indiana 47405, USA}
\affil[9]{Department of Physics, Yale University, New Haven, CT USA}
\affil[10]{Department of Physics, University of Warwick, Coventry, CV4 7AL, UK}
\affil[11]{Physics Department, Boston University, Boston, USA}
\affil[12]{Uninettuno University, Rome, Italy}
\affil[13]{INFN, Sezione di Roma Tor Vergata, Rome, Italy}
\affil[14]{INAF, Osservatorio Astronomico di Roma, Monteporzio Catone, Italy}
\affil[15]{Physics Department, Indiana University, Bloomington, IN 47405, USA}
\affil[16]{Department of Mechanical Engineering, Massachusetts Institute of Technology, Cambridge, MA 02139, USA}

\maketitle

\begin{Abstract}
\noindent 
\small
High-energy physics is primarily concerned with uncovering the laws and principles 
that govern nature at the fundamental level.
Research in this field 
usually relies on probing the boundaries of established physics,
an undertaking typically associated with extreme energy and distance scales.  
It is therefore unsurprising
that particle physics
has traditionally been dominated by large-scale experimental methods  
often involving high energies, 
such as colliders and storage rings, 
cosmological and astrophysical observations,
large-volume detector systems, etc.
The corresponding measurements
are ideally suited 
for the discovery of new particles and interactions.

However, 
high-sensitivity measurements in smaller experiments,
often performed at lower energies, 
are presently experiencing a surge in importance 
for particle physics for at least two reasons.
First, 
they exploit synergies to adjacent areas of physics
with recent advances in experimental techniques and technology. 
Together with intensified phenomenological explorations, 
these advances have led to the realization that
challenges associated with weak couplings 
or the expected suppression factors for new physics
can be overcome with such methods
while maintaining a large degree of experimental control. 
Second, 
many of these measurements 
broaden the range of particle-physics phenomena and observables 
relative to the above set of more conventional methodologies. 
Combining such measurements 
with the conventional efforts above
therefore casts both a wider and tighter net
for possible effects 
originating from physics beyond the Standard Model (BSM).

The present work argues 
that this assessment
points at a growing impact of such methods and measurements 
on high-energy physics,
and it therefore warrants direct support 
as particle-physics research. 
More specifically,
we discuss a sample of ongoing and future efforts 
in this context 
involving 
cold neutrons,
a range of AMO-based studies,
first- and higher-generation antimatter, 
and microscopic mechanical experiments
including gravitationally entangled masses 
and optically levitated nanospheres.
These efforts are poised 
to yield crucial insights into proposed BSM physics 
as diverse as
novel short-range interactions,
the small-scale structure of spacetime 
and in particular the fate of Lorentz, translation, CPT, CP, T, and P symmetries, 
the gravitational interaction of antimatter,
certain quantum aspects of gravity,
millicharged particles,
gravitational-wave measurements,
and dark matter.
These synergies and their prospective physics output
foreshadow a promising future
for such types of experimental and theoretical activities. 
Leveraging the recent rapid progress and bright outlook 
associated with such studies
for high-energy physics,
could yield high returns,
but requires substantial and sustained efforts by funding agencies. 
\normalsize
\end{Abstract}

\tableofcontents

\def\thefootnote{\fnsymbol{footnote}}
\setcounter{footnote}{0}

\section{Introduction}

High-energy physics is broadly aimed 
at exploring the world  
at extreme scales 
and routinely involves the search 
for novel interactions and degrees of freedom.
General effective-field-theory considerations suggest
that the sizes of the corresponding new-physics effects
can be organized into powers of $\kappa\frac{E}{M}$,
where $E$ is the energy scale of the process under consideration,
$M$ characterizes the energy scale of the underlying physics, 
and $\kappa$ is an appropriate coupling constant. 
In light of the expected large size of $M$,
the traditional approach to experimental progress in this field 
is
high-energy measurements at colliders and in astrophysics.
Experimental explorations involving high-intensity physics 
represent an additional pathway forward,
in particular in situations with small couplings~$\kappa$.

The recent development of low-energy ultrahigh-precision physics techniques and ideas
has opened a further, complementary avenue 
to overcome the typical high-suppression factors
in the search for underlying physics: 
low-energy small-scale experiments.
They represent versatile experimental tools 
for such purposes,
and the primary goal of this article 
is to expose their suitability
for examining the foundations of physics,
such as spacetime symmetries, 
gravitational physics, 
quantum mechanics, 
and their interplay. 
The paragraphs below
provide an overview of ideas for such studies within the next decade.
More detailed descriptions of these anticipated activities 
are contained 
in
the subsequent sections.

Spacetime symmetries underlie many features 
in a wide variety of physical systems 
and can therefore be investigated 
with a correspondingly broad range of experimental techniques. 
One set of possible studies in this context 
concerns searches for T- and P-odd interactions  
in slow-neutron--nucleus reactions.
Boosted by neutron-nucleus resonances,
the physics reach of such studies 
is comparable to that of neutron and nuclear EDM searches.  
We note that this white paper does not provide a detailed description of physics opportunities with searches for electric dipole moments of fundamental particles. Such physics opportunities are described in detail in another white paper submitted to the Snowmass proceedings \cite{Alarcon:2022ero}.

Another set of spacetime-symmetry investigations in the laboratory
is the search for violations of translation invariance 
through varying fundamental couplings. 
The spectroscopy of a carefully chosen nuclear transition in the $^{228}$Th nucleus 
represents a promising idea in this context. 
Such a measurement has the potential 
to surpass the $10^{-18}$ precision 
of state-of-the-art atomic clocks 
by two to three orders of magnitude,
and would provide an ultra-sensitive probe
for the constancy of electromagnetic and other couplings. 
Likewise, 
the aforementioned slow-neutron resonance interactions 
can also be employed to produce the most sensitive laboratory constraints 
on possible time dependences of dimensionless parameters in QCD,
such as $m/\Lambda _{\rm QCD}$, 
where $m$ is the scale of the light quark masses.

Lorentz- and CPT-invariance tests 
represent a third set of investigations
in this context (Sec.~\ref{CPT}). 
Various theoretical approaches to physics beyond the Standard Model and general relativity 
are known to accommodate tiny departures from these closely intertwined symmetries. 
This insight has spawned phenomenological studies in effective field theory 
that have identified many
potential
signatures of such symmetry violations 
in low-energy small-scale experiments as diverse as
antihydrogen spectroscopy and free-fall studies, 
clock-comparison tests, 
cold-neutron measurements, 
matter-wave interferometry, 
muon physics, 
Penning-trap tests, 
resonant-cavity measurements, 
and short-range interaction studies.
Such experimental efforts are ongoing with ample territory still to cover,
and in some cases they have already reached Planck sensitivity.

The unique theoretical structure of the gravitational interaction
including its resistance to a quantum description 
as well as the experimental challenges posed by its weakness 
make gravity an interesting candidate as a harbinger of new physics. 
A key unresolved question in this arena is
whether gravity can mediate quantum entanglement. 
This question can be addressed experimentally 
in a low-energy small-scale study
utilizing the toolbox of quantum optomechanics. 
The idea is to measure and prepare quantum states of motion 
of two neighboring masses 
and monitor their time evolution
\!\!,
with decoherence being a tell-tale signature 
of gravity not fully obeying the laws of quantum mechanics.

A second active research area involving the gravitational interaction
concerns its short-range behavior.
Corrections to the Newtonian inverse-square law 
are predicted in a number of models for new physics 
spurring precision measurements of gravity 
at sub-millimeter distance scales. 
A promising experimental approach to such measurements
is
optically-levitated dielectric nanospheres in high vacuum
\!\!,
because they achieve excellent decoupling from their environment 
and allow force sensing at the zeptonewton level ($10^{-21}$N). 
Such a system would
allow tests of the inverse-square law at the micron scale,
and it would also permit a range of other fundamental-physics measurements including
searches for gravitational waves, millicharged particles, and Dark Matter,
as well as studies of the aforementioned role of gravity in quantum entanglement. 
Slow-neutron interferometry 
provides a complementary precision probe for similar novel interaction effects.
Examples include sensitivities to 
exotic short-range gravity ($10^{-8}$\,m to $10^{-13}$\,m)
and novel, weakly coupled spin-dependent interactions ($10^{-3}$\,m to $10^{-8}$\,m).

A third class of gravity investigations 
involves the gravitational interaction with antimatter and other exotic matter. 
These interactions may, 
for example, 
be modified in the presence of CPT and Lorentz violation, 
but they are just beginning to be explored experimentally. 
The neutral-kaon
system provides unique access to such effects: 
a putative difference in the gravitational interaction 
between its matter and antimatter components 
would lead to measurable effects in their oscillation pattern (Sec.~\ref{section-kaon-system}).
Muonium interferometry offers another promising avenue 
for 
the pursuit of
such investigations. 
Exploratory studies 
by the MAGE collaboration
involving a novel muonium beam under development 
support the feasibility 
to determine the terrestrial gravitational acceleration of antimatter 
at the percent level (Sec.~\ref{section-mage}).
This idea could also lead to the first gravitational measurements of purely leptonic matter 
and of 2nd-generation matter.

\section{Tests of Spacetime Symmetries}

\subsection{NOPTREX: A Neutron OPtics Time Reversal EXperiment to search for Time Reversal Violation in Neutron-Nucleus Resonance Interactions}
\label{NOPTREX}

\subsubsection{Introduction}

New sources of time reversal violation are needed to explain the baryon asymmetry of the universe in Big Bang cosmology according to the Sakhaorv argument~\cite{Sak67}. Neutron interactions with heavy nuclei at certain compound nuclear p-wave resonances can be used to search for P-odd/T-odd interactions through a term in the neutron forward scattering amplitude of the form $\vec{s}_{n} \cdot (\vec{k}_n \times \vec{I})$, where $\vec{s}_{n}$ is the spin of the neutron, $\vec{k}_{n}$ is the neutron momentum, and $\vec{I}$ is the spin of the nucleus. The highly excited states in heavy nuclei involved in this type of search offer a qualitatively different environment from the ground states probed by electric dipole moment experiments of nucleons and nuclei. The ratio of the P-odd and T-odd amplitude to the P-odd amplitude on the same p-wave resonance is quite insensitive to unknown properties of the compound resonant states involved. In the case of the forward elastic neutron scattering amplitude, since the state of the polarized target does not change and since the optical theorem relates the imaginary part of the forward scattering amplitude to the cross section, the cross section differences for the forward and time reversed processes are proportional to amplitude differences and therefore can realize a sensitive null test for T invariance which is in principle free from the effects of final state interactions~\cite{Gudkov:1990tb,Gudkov:2013dp,Bowman:2014fca}.

\subsubsection{Experimental Approach}

Amplifications of $P$-odd neutron amplitudes in compound nuclear resonances by factors of $~10^6$ above the $~10^{-7}$ effects expected for weak NN amplitudes compared to strong NN amplitudes have already been observed~\cite{Mitchell:1999zz} in measurements of $\Delta \sigma_{P}$ several heavy nuclei, including some at $p$-wave resonances in the few $eV$ energy range such as $^{139}\mathrm{La}$~\cite{Yuan:1991zz}, $^{131}\mathrm{Xe}$~\cite{Szymanski, Skoy}, and $^{81}\mathrm{Br}$~\cite{Dubna, Frankle, Shimizu}, and $^{117}$Sn. This amplification from mixing of nearby s and p-wave resonances was predicted theoretically before it was measured, and the same resonance amplification factor applies to a $P$-odd and $T$-odd amplitude up to factors of order unity. Although the nuclear states involved are extremely complicated at the level of the many-body nuclear wave functions, one can form a dimensionless ratio $\lambda_{PT}= \frac{\Delta \sigma_{TP}}{\Delta \sigma_{P}}=\kappa(J) \frac{<\phi_{p}|V_{PT}|\phi_{s}>}{<\phi_{p}|V_{P}|\phi_{s}>}$ of the T-odd, P-odd asymmetry $\Delta \sigma_{TP}$ of interest to the measured P-odd asymmetry $\Delta \sigma_{P}$ at the position of the enhanced p-wave resonance energy, the ratio $\frac{<\phi_{p}|V_{PT}|\phi_{s}>}{<\phi_{p}|V_{P}|\phi_{s}>}$ of the matrix elements of the P-odd and T-odd interaction to the P-odd interaction between the same pair of s and p wave resonance states $|\phi_{s}>$ and $|\phi_{p}>$, and a spin-weighted sum of resonance partial widths $\kappa(J)$ which can be determined experimentally using (n, $\gamma$) spectroscopy. Since this ratio involves expectation values in the same compound nuclear wave functions it can possess a clean theoretical interpretation. Similar considerations apply also to P-even and T-odd interactions: they can also generate a term in the neutron forward scattering amplitude which possesses resonant amplification. 

The statistical uncertainty that could be achieved in such an experiment after $10^7$ seconds of data in $^{139}$La at a MW-class short pulse neutron spallation source implies that one can measure the ratio $\lambda_{PT}$ to $1 \times 10^{-4} - 1 \times 10^{-5}$ sensitivity, which translates into an improved sensitivity to P-odd and T-odd neutron-nucleus interactions of about an order of magnitude~\cite{Bunakov:1982is,Gudkov:1991qg,Skoy:2007,Gudkov:2013dp,Bowman:2014fca}. The $0.7\ eV$ resonance in $^{139}$La has a P-odd longitudinal asymmetry of $9.5\%$~\cite{Yuan:1991zz} and is therefore a good candidate for this search. $\kappa$ has been constrained recently in $^{139}$La~\cite{Okudaira2018} to be at least of order 1, and ongoing experiments at JPARC will soon measure $\kappa$ in other NOPTREX candidate nuclei. Groups at KEK~\cite{Masuda_90}, Kyoto University~\cite{Takahashi:1993np}, and PSI~\cite{Hautle:2000} achieved substantial (up to 50\%) polarization of $^{139}$La nuclei in lanthanum aluminate crystals in volumes as large as $~10\ cc$, enough for the experiment, and R\&D to polarize $^{81}$Br~\cite{Keith} and $^{131}$Xe and $^{117}$Sn~\cite{Goodson} is in progress. Ongoing R\&D on high phase space acceptance supermirror neutron optics has the potential to improve the statistical sensitivity in the future by another order of magnitude.

The bright pulsed sources of epithermal neutrons at MW-class spallation neutron facilities like SNS and JSNS have enough intensity at eV energies to reach the statistical accuracy required for a sensitive search. The separation of neutron energies by time-of-flight from these pulsed sources also allows a powerful search for systematic errors by looking above and below the neutron resonance energy at both the transmitted and scattered neutrons. Existing technology for eV neutron polarization using polarized $^{3}$He neutron spin filters suffice for the measurement.

\subsubsection{Conclusions}

This estimated sensitivity accessible today is comparable to that being proposed for the next-stage neutron EDM searches. However as the neutron-nucleus system possesses interactions not present in the single neutron system involved in nEDM searches, it is quite possible that P and T violation might be seen in one of these observables but not the other~\cite{Pospelov:2001ys,Pospelov:2005pr,Song:2011jh,Gudkov:2013edm}. In particular, the NOPTREX observable is sensitive to axion-like particles with masses in the eV-MeV range~\cite{Mantry2014,Fadeev2019}. It is therefore very important to pursue such a search if one can suppress the potential sources of systematic error. As no such polarized neutron optics search for P-odd and T-odd interactions has ever been conducted, the first real experiment will represent a pioneering effort.   

Birefringent neutron optical devices recently developed for neutron spectroscopy can convert the NOPTREX experimental apparatus into a spin-path interferometer, similar to the Ramsey separated oscillatory field configuration used in electric dipole moment searches but operating with paths separated in space rather than in time. These devices were recently used to entangle the neutron spin and position or the neutron spin, position, and energy variables into Bell and GHZ states, whose degree of quantum entanglement was quantified by measuring the appropriate Bell and GHZ entanglement witnesses~\cite{Shen2020,Lu2020}. The correlation observables in this experiment took the largest possible value allowed by quantum mechanics despite the passage of the polarized neutrons through macroscopic amounts of matter. The small decoherence of the transmitted neutron state confirmed by this work implies that neutron interferometric methods based on this technology can be applied to NOPTREX to help isolate the P-odd/T-odd signal of interest from many possible sources of systematic error and help ensure that the neutron optical T-odd null test condition is satisfied.

\subsection{Lorentz and CPT Tests with Low-Energy Precision Experiments}
\label{CPT}
 
 \subsubsection{Introduction}

The role of Lorentz symmetry in physics 
can hardly be overstated.
When combined with quantum mechanics 
and a few mild physical assumptions,
it yields relativistic quantum field theory~\cite{Weinberg:1995mt}
together with a further symmetry, 
CPT invariance~\cite{Streater:1989vi}.
This framework
constitutes the basis for the Standard Model, 
which is our best description of nongravitational physics.
In addition, 
Lorentz and CPT symmetry 
are typically a key ingredient in theoretical explorations 
of physics beyond the Standard Model.

The extraordinary relevance of these spacetime symmetries alone
provides abundant motivation 
for their continued experimental and theoretical study.
Further significant impetus for improved Lorentz and CPT tests 
derives from a number of BSM physics ideas.
Despite being based on these symmetries,
they allow for small departures from Lorentz and CPT invariance
in the ground state 
with signatures accessible with current and near-future technology.
Examples include spontaneous CPT and Lorentz breaking in string theory,
through noncommutative field theory, 
and through cosmologically varying scalars~\cite{Kostelecky:1988zi,
	Kostelecky:1991ak,
	Mocioiu:2000ip,
	Carroll:2001ws,
	Carlson:2001sw,
	Anisimov:2001zc,
	Alfaro:1999wd,
	Kostelecky:2002ca,
	Arkani-Hamed:2003pdi,
	Jackiw:2003pm}.

For the identification, interpretation, and comparison 
of Lorentz and CPT tests 
in a largely model-independent way 
a general framework 
called the Standard-Model Extension (SME)~\cite{Colladay:1998fq,Kostelecky:2003fs,Kostelecky:2009zp,Kostelecky:2011gq,Kostelecky:2013rta,Kostelecky:2018yfa}
has been developed.
The SME is based on effective field theory
and incorporates both the usual Standard Model and General Relativity as limiting cases,
and over the last two decades it has matured into the standard phenomenological tool
for Lorentz- and CPT-violation searches 
in the entire body of established physics.
With hundreds of past experimental constraints on Lorentz and CPT violation~\cite{Kostelecky:2008ts},  
this topic has been on a climbing trajectory 
and is poised to gain further momentum in the coming decade.
The next section contains brief descriptions of small-size low-energy physical systems 
with demonstrated impact on the field 
and substantial future promise for record sensitivities.


\subsubsection{Experimental approaches}

{\bf Antihydrogen measurements.}
The availability of cold antiprotons at CERN's Antiproton Decelerator
has paved the way for unprecedented studies of antihydrogen. 
One class of these is concerned with antihydrogen precision spectroscopy:
the ALPHA and ASACUSA 
experiments 
are designed for such antihydrogen measurements, 
including 1S--2S, 1S--2P, and hyperfine spectroscopy,
and compare these to the corresponding frequencies in ordinary hydrogen 
for a direct CPT test~\cite{AntiHBook}. 
These efforts are well underway 
with the completion of various extraordinary milestones,
such as a 1S--2S measurement 
just three orders of magnitude 
shy of the corresponding accuracy in hydrogen. 
Interpreted in terms particle--antiparticle absolute mass differences,
this measurement exceeds, 
for the first time,
the precision attained in neutral-kaon interferometry,
a system considered the particle-physics standard for CPT tests~\cite{ALPHA1,ALPHA2,ALPHA3,ALPHA4,ALPHA5,ASACUSA1,ASACUSA2}.
Another class of antihydrogen experiments seeks to study 
the interaction of antimatter with gravity. 
For example, 
AEgIS, ALPHA-g, and GBAR at CERN will be employing complementary methods 
to measure the rate of free fall of antihydrogen 
in the gravitational field~\cite{AntiHBook}, 
and a proposal for a further antimatter gravity experiment at Fermilab exists~\cite{AGE}.
Both spectroscopic and free-fall efforts 
are currently straining at the leash 
to resume antihydrogen studies
as the current Long Shutdown~2 
at the LHC
draws to a close
and the new Extra-Low Energy Antiproton Ring ELENA 
goes into full operation.
The community will then be within striking distance 
for qualitatively novel Lorentz and CPT tests 
within effective field theory.


\vskip+6pt
{\bf Comparative studies of protons and antiprotons in Penning traps.}
Penning traps permit the isolation and investigation of
individual charged particles and antiparticles.
Lorentz and CPT tests with such devices 
are typically based on two types of measurements:
sidereal time variations in the cyclotron and anomaly frequencies of trapped particles 
as the Earth rotates about its axis 
and instantaneous anomaly-frequency comparisons 
between particles and antiparticles.
Numerous past studies have contributed to
bounds on Lorentz and CPT violation 
that can be considered as probing the Planck regime~\cite{bluhm-97,bluhm-98,dehmelt-99,mittleman-99,gabrielse-99,fitante12,ulmer15,ding16,nagahama17,smorra17,smorra19,BASE:2022yvh}.
Efforts in this field are bound to gain even further momentum in the future.
For example, 
prospective upgrades at the BASE experiment,
such as quantum-logic based spin readout~\cite{qlr} 
a portable antiproton trap~\cite{Portable}
as well as recent phenomenological progress~\cite{Kostelecky:2021tdf} 
paving the way for studies of the gravitational interaction of antimatter in penning traps
will allow access to a much enlarged set of Lorentz- and CPT-breaking observables 
as well as substantial gains in sensitivity. 


\vskip+6pt
{\bf Clock comparisons.}
Some of the sharpest Lorentz-violation bounds for protons, neutrons, electrons, and photons, 
which can reach sensitivities of up to $10^{-29}$ for certain types of light-speed anisotropies,
stem from atomic clocks, atom magnetometry, and other precision spectroscopy experiments~\cite{Kozlov:2018mbp,MB19,SH19,FR17,PG17,PR15,AH14,HL13,MP13,SB11,Cane:2003wp,Bear:2000cd}.
Clock comparisons involve performing high-precision comparative measurements 
of at least two transitions in atomic clocks 
as the Earth rotates: 
anisotropies arising from violation of Lorentz symmetry 
are predicted to produce orientation dependence 
in the difference between the two clock frequencies~\cite{clockKV1,clockKV2}.
On the other hand, 
clock-comparison experiments performed in space 
aboard an orbiting platform,
such as the International Space Station, 
with a laboratory frame 
that is both rotating and boosted
provide sensitivities to forms of Lorentz breaking
that are not readily testable in terrestrial laboratories~\cite{Bluhm:2003un}. 
The last decade has witnessed remarkable improvements 
in optical clocks and trapped-ion control
that were utilized for numerous Lorentz-symmetry tests 
with extraordinary precision~\cite{MB19,SH19,PR15,HL13}.
In the future, 
this trend is expected to pick up pace
with novel measurement schemes specifically designed to improve clock comparisons 
by orders of magnitude~\cite{Shaniv:2017gad} 
and rapid improvements in clock precision 
and the development of new clock technologies~\cite{Kozlov:2018mbp}.


\vskip+6pt
{\bf Cold neutrons.}
Due its unique combination of physical properties, 
such as neutrality, small Compton wavelength, low polarizability, 
and high matter-penetration power, 
the neutron has long been employed as an indispensable tool in experimental research 
including Lorentz and CPT tests. 
For example, 
ultrahigh sensitivities to SME coefficients have been attained 
via measurements involving 
neutron-spin motion~\cite{Altarev:2009wd},
neutron--antineutron oscillations~\cite{Babu:2015axa},
and gravitationally bound neutrons~\cite{Ivanov:2019ouz}.
With various prospective nEDM measurements
at different laboratories, 
such as PSI~\cite{PSI_EDM}, 
ILL~\cite{Wurm:2019yfj},
TRIUMF~\cite{TRIUMF_EDM},
and SNS~\cite{SNS_EDM},
current constraints on neutron SME coefficients 
can be improved by up to about two orders of magnitude, 
and previously unexplored SME observables 
can be measured. 
Likewise,
the planned NNbar experiment at ESS 
will provide unprecedented sensitivity 
to neutron--antineutron oscillations~\cite{NNBAR}.


\vskip+6pt
{\bf Matter-wave interferometry.}
Lorentz breakdown can also deform 
the interaction of gravity with matter~\cite{akgrav,bailey2006,tasson2009,tasson2011}.
The ensuing physical effect can therefore be explored 
with experimental techniques
such as superconducting gravimeters and space-based missions~\cite{tasson2017,shao2018,pihan2019},
which continue to increase in sensitivity, 
and proposals for gravitational measurements with exotic systems, 
such as ones involving antimatter or higher generations~\cite{muonGKV,muonium,Antognini:2018nhb}, 
exist.
Gravitational phenomena 
are also amenable to studies with matter-wave interferometers~\cite{dimopoulos2008}
and have already placed bounds on Lorentz violation
when used as gravimeters~\cite{mueller2008} 
and as equivalence-principle tests~\cite{hohensee2011}.
Future atom-interferometer methods are expected to compete 
with these recent advances~\cite{dimopoulos2008,Asenbaum:2020era}.
In particular, 
capabilities such as large wave-packet separation
in both space and momentum~\cite{space_sep,mom_sep}
as well as simultaneous multispecies operation~\cite{Asenbaum:2020era,Schlippert:2014xla,Hartwig:2015iza}, 
promise leaps in both sensitivity and versatility 
of SME tests~\cite{Schlippert:2019hzx}. 
Extrapolating such developments, 
matter interferometry 
will be positioned at the forefront 
of probing Lorentz symmetry
at the interface of matter and gravity 
in the coming years.


\vskip+6pt
{\bf Muon physics.} 
The history of Lorentz tests involving muons 
dates back almost 80 years 
to a measurement establishing relativistic time dilation. 
At present, 
muon systems are again scrutinized for new physics 
including Lorentz and CPT breakdown~\cite{BKLmuon,muonGKV}. 
One of these systems is muonium: 
its theoretical tractability 
and experimental accessibility 
have stimulated clean spectroscopic Lorentz and CPT tests 
with unique sensitivities to SME coefficients~\cite{Muonium01}.
The future ground-state hyperfine spectroscopy 
by MuSEUM at J-PARC~\cite{MuSEUM},
the proposed determination of the 1S--2S transition frequency 
by Mu-MASS at PSI~\cite{Mu-Mass}, 
and proposals for gravity measurements with muonium~\cite{muonium,Antognini:2018nhb}
are clear indications for the growing vitality of the field 
in the coming years. 
Muon-spin precession represents a further experimental avenue
in this context 
because spin motion is affected 
by various SME coefficients.
This idea has already provided the basis 
for past analyses of muon $g-2$ data~\cite{BKLmuon,muong208,muonGKV}.
Future studies of $\mu^+$ spin motion,
such as Muon $g-2$ at Fermilab~\cite{muong215}
and E34 at J-PARC~\cite{muong219},
are in an exquisite position 
to sharpen existing Lorentz and CPT tests 
and access unconstrained SME observables~\cite{muong219F}. 
An additional $\mu^{-}$ run at the Fermilab experiment 
would permit a direct CPT test,
further broadening the scope of such efforts. 


\vskip+6pt
{\bf Resonant cavities.}
Lorentz tests with
electromagnetic resonant cavities
are modern versions of the classic
Michelson--Morley experiment~\cite{MichelsonMorley1,MichelsonMorley2}
and provide high sensitivities to the photon's SME coefficients.
They typically compare the resonant frequencies
of two cavities at different orientations
and look for variations as the cavities are rotated or boosted.
To date, experiments utilizing
microwave cavities~\cite{microwave1,microwave2,microwave3,microwave4,microwave5,microwave6,nonminmicrowave},
optical cavities~\cite{optical1a,optical1b,optical1c,optical2a,optical2b,optical2c,optical2d},
ring resonators~\cite{rings1a,rings1b,rings2,nonminrings1,nonminrings2}, 
and acoustic cavities~\cite{acoustic1,acoustic2}
have placed tight constraints on deviations
from perfect Lorentz invariance.
The LIGO interferometer has also been used to perform
a more traditional Michelson--Morley experiment~\cite{ligo}.
The last two decades have seen sensitivities in cavity experiments 
improve by orders of magnitude
and an ever expanding reach into different forms
of Lorentz violation~\cite{nonminmicrowave,nonminrings1,nonminrings2}.
This trend is expected to continue in future experiments,
including those performed in space~\cite{mstar}.


\vskip+6pt
{\bf Short-range-interaction studies.}
Precision measurements set up to probe the gravitational inverse-square law 
and search for novel interactions 
typically exhibit intrinsic geometrical orientations, 
such as specific arrangements of test bodies.
This feature makes them also ideal candidates for Lorentz and CPT tests: 
laboratory motion, 
such as sidereal revolution about the Earth's axis,
typically changes this orientation, 
opening the possibility to detect fundamental anisotropies 
in the physics of the system under investigation~\cite{SRG_Theo}.
This idea has produced some of the best experimental constraints 
on the SME's gravity sector~\cite{SRG_Exp1,SRG_Exp2}, 
and planned experimental upgrades~\cite{Chen:2017bru}
provide further impetus for future efforts along these lines. 
An additional idea in this context 
concerns experiments with a spin-polarized torsion pendulum~\cite{bluhm-00a}.
The corresponding  measurements have placed stringent limits 
on spatial-anisotropy coefficients~\cite{ni-03,heckel-06,heckel-08},
and the ongoing improvement of such methods~\cite{Luo:2020gjh,Zhu:2018mrf,Lee:2020zjt} 
bodes well for continued activity in this field in the coming decade.

\subsubsection{Conclusion}

Lorentz and CPT symmetry 
are foundational principles
within the boundaries of established high-energy physics 
as well as key assumptions in most theoretical approaches 
to expand those boundaries.
At the same time,
a number of these theoretical approaches 
allow for ground states 
exhibiting small departures from these symmetries. 
In light of this dual significance, 
the continued scrutiny of Lorentz and CPT invariance 
assumes particular urgency 
in particle physics.
Present-day and near-future experimental efforts
are on track to deliver low-energy high-precision Lorentz and CPT tests
with the distinct potential 
to uncover qualitatively new physics
with Planck-scale reach.
Phenomenological and experimental Lorentz- and CPT-symmetry studies
therefore fall within the confines of high-energy physics,
are critical to the future of the community,
and should be intensified.

\section{Tests of fundamental symmetries related to gravity}

The question of antimatter gravity, first raised in the 1950s~\cite{Morrison:1958}, is of continuing interest~\cite{Nieto:1991,Fischler:2008}. In the 
``antigravity'' scenario, antimatter is predicted to repel matter~\cite{Morrison:1958,Kowitt:1996,Chardin:1996,Banchet:2008,Burinskii:2008,Blanchet:2009,Cabbolet:2010,Hajdukovic:2011,villata:2011,benoitlevy:2012,Hajdukovic:2012,Villata:2012,Villata:2013,Villata:2015}. 
This is well motivated, since a universe comprising equal amounts of matter and antimatter that repel  gravitationally could 
(i) explain the missing antimatter, 
(ii) fit supernova data without dark energy~\cite{benoitlevy:2012,DiracMilne2014}, and
(iii) explain galactic rotation curves with gravitational vacuum polarization rather than with dark matter~\cite{Banchet:2008}.
Such a universe would also (iv) have expanded slowly enough to
explain the uniform temperature of the cosmic microwave background radiation without cosmic inflation~\cite{benoitlevy:2012,DiracMilne2014}.
On the other hand, in a field-theory-motivated framework, the gravitational acceleration of antimatter by matter might differ only slightly from that of matter~\cite{Nieto:1991}, contrary to expectations from general relativity, and perhaps provide clues to the correct quantum theory of gravity. 
Decades of experimental effort
have yet to yield a statistically significant direct measurement. Antimatter gravity studies using antihydrogen (${\overline {\rm H}}$) are ongoing~\cite{osti_1623908,AEGISProto:2008qxw,GBAR:2011fij}, and experiments with positronium have been discussed~\cite{Cassidy:2014IJMPS..3060259C}. Here, we discuss a possible direct measurement using muonium (Section~\ref{section-mage}) and a novel indirect measurement using neutral kaons (Section~\ref{section-kaon-system}).

\subsection{Muonium Antimatter Gravity Experiment (MAGE)} 
\label{section-mage}

\subsubsection{Introduction}\vspace{-.05in}
\label{intro}
\global\long\def\gbar{\ensuremath{\bar{g}}}
 
We here consider a possible measurement with muonium (M or Mu), an~exotic atom consisting of an electron bound to an antimuon; unlike the ${\overline {\rm H}}$ case, the interpretation of such a measurement  has no hadronic uncertainties. This measurement\,---\,the goal of the Muonium Antimatter Gravity Experiment (MAGE) collaboration\,---\,could potentially be performed at an upgraded Fermilab muon complex~\cite{LEmuon-LoI}.

The most sensitive ($\sim10^{-7}$) limits on antimatter gravity  come from {\em indirect} tests  (for example, equivalence principle tests using torsion pendula~\cite{Wagner:2012} or masses in Earth orbit~\cite{PhysRevLett.119.231101}), relying on the expected amounts of virtual antimatter in the 
atoms of various elements~\cite{alves2009experimental}; these are invalid in the antigravity scenario and, in any case, are inapplicable to muonium. Another limit,  $|\alpha_g-1|<8.7\times10^{-7}$~\cite{ulmer15}, has been derived from the measured cyclotron frequency of magnetically confined antiprotons, compared with that of H$^-$ ions, based on the gravitational redshift due to Earth's gravitational potential in the field of the local galactic supercluster~\cite{Hughes:osti_5751101,PhysRevLett.65.1317,good-1961}; it too need not apply to antimuons.\footnote{And we note that arguments based on absolute gravitational potentials have been critiqued by Nieto and Goldman~\cite{Nieto:1991}. Other precise measurements of these cyclotron frequencies~\cite{gabrielse-99,smorra17} have not been interpreted in terms of possible matter--antimatter gravitational differences.} 

A {\em direct} test of the gravitational interaction of antimatter with matter is desirable on quite general grounds~\cite{Nieto:1991}.\footnote{The only published direct test so far~\cite{osti_1623908} has yielded the limit  $-65 < \gbar/g < 110$.} Such a measurement can be viewed as a test of general relativity
or as a search for a fifth force and is of interest from both perspectives. Recent work~\cite{tasson2011,tasson2015gravity,tasson2016symmetry} on the SME emphasizes the importance of second-generation gravitational measurements. Current interest in “fifth force” models~\cite{GlashowPhysRevLett.114.091801,Buttazzo2017} (stimulated by evident anomalies in the leptonic decays of $B$ mesons) also supports more detailed investigations of muonium.

\vspace{-.1in}\subsubsection{Experiment Concept}\vspace{-.05in}

\begin{figure}[tb]
  \begin{minipage}[c]{0.42\textwidth}
    \caption{
MAGE experiment concept (elevation view; gravitational deflection and phase shift $\Delta\phi$ exaggerated for clarity). Muonium beam enters from left, slow-electron detector is at right. Not shown: ring electrodes to accelerate slow electrons onto their detector, starting downstream of grating 3 and continuing within scintillating-fiber-barrel positron detector;  hodoscope around positron barrel.}\label{fig:MAGE-concept}
  \end{minipage}\vspace{-.2in}
  \begin{minipage}[c]{0.57\textwidth}\vspace{-.15in}
    \includegraphics[width=\textwidth]{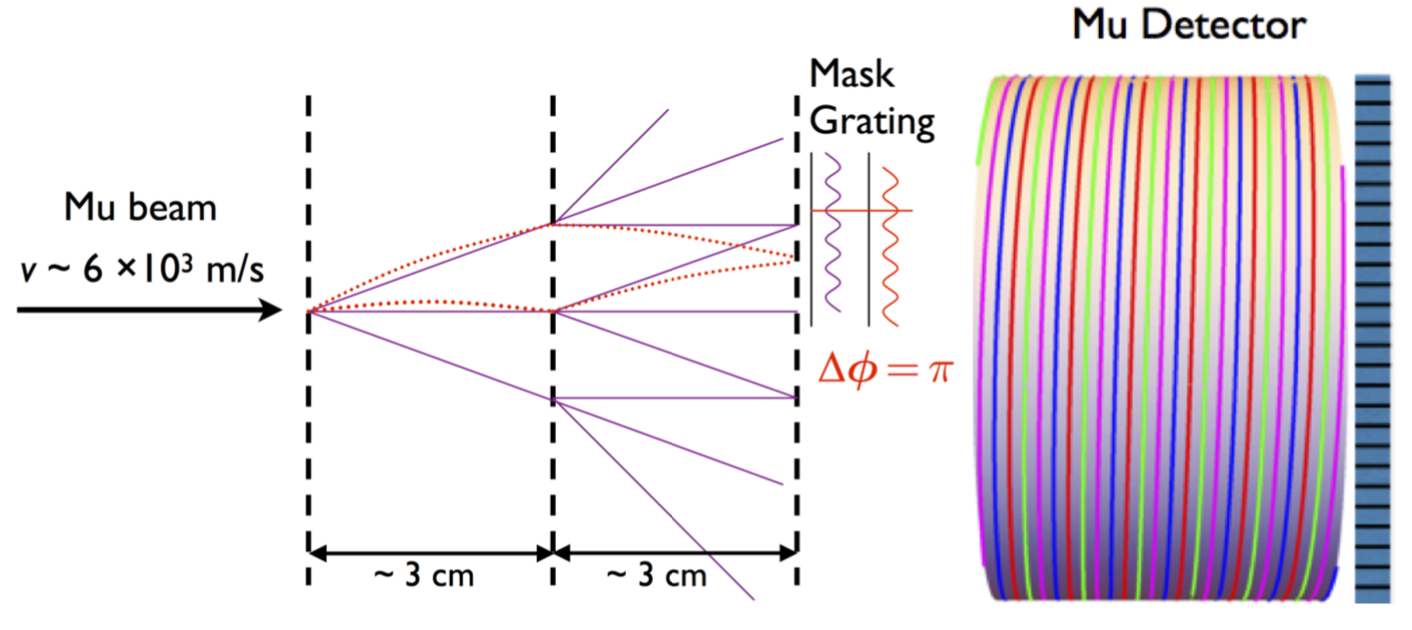}
  \end{minipage}
\end{figure}

A direct test of antimatter gravity can be performed interferometrically, by passing an intense, high-quality muonium beam in vacuum through precise nanofabricated gratings and measuring the gravity-induced phase shift~\cite{Antognini:2018nhb,Kirch:2014,kirch2007testing}.  As shown in Fig.~\ref{fig:MAGE-concept},  a horizontal, parallel, slow muonium beam impinges on  a 3-grating, Mach--Zehnder-type interferometer, with the interference pattern following the beam's gravitational  acceleration. Mu atoms  decaying after the third grating are detected as a coincidence between a fast positron in the barrel detector and a slow electron electrostatically accelerated onto a microchannel plate at the back. The interferometric phase is measured by translating a grating continually up and down and analyzing the resulting changes in detected coincidence rate. The phase is quite small: $\Delta\phi = 2\pi \gbar t^2/d\approx 0.01$ (for $\gbar = g$), where $t$ is the time for the atom to traverse the distance between gratings and $d$ is the grating pitch (here taken as 100\,nm). The required few-picometer alignment system is feasible using laser interferometry~\cite{Antognini:2018nhb,Kaplan:2018,Thapa:2011}. The zero-deflection phase is determined by periodically illuminating the interferometer with soft X-rays, with a systematic check provided  by periodically rotating the interferometer by 90 or 180$^\circ$.

Preparing the intense, high-quality Mu beam needed for MAGE is a challenge. Beam R\&D is currently carried out at Switzerland's Paul Scherrer Institute (PSI)~\cite{PhysRevLett.125.164802,Iwai:2020jye} following ideas of Taqqu~\cite{TAQQU2011216,Taqqu:PhysRevLett.97.194801}, involving cooling of a surface muon beam in  gaseous helium in crossed electric and magnetic fields to reduce its 6D emittance by some 10 orders of magnitude, at a cost of two to three orders of magnitude in muon decay loss. The cooled beam can then be stopped in a $\sim\mu$m-thick layer of superfluid helium (SFHe) at the bottom of a cryostat, efficiently forming muonium, which is then expelled vertically from the upper SFHe surface at a predicted speed of 6.3\,mm/$\mu$s~\cite{Taqqu:PhysRevLett.97.194801} due to its expected large, positive chemical potential (270\,K) in SFHe~\cite{TAQQU2011216}. The vertical beam is turned to  horizontal, as needed for MAGE, by means of a 45$^\circ$\ SFHe-coated deflector~\cite{Luppov:PhysRevLett.71.2405}. (Because the Mu
atoms are in thermal equilibrium with the SFHe prior to expulsion, both the  beam energy spread
and its angular divergence  are determined by the ratio of the $\sim 0.2$\,K SFHe temperature to the Mu chemical potential.) The resulting interferometer acceptance is maximal,
leading to a  $5\sigma$ \gbar\ sign determination  with about one month's worth of beam at PSI~\cite{Antognini:2018nhb}.

Another beam option exploits  another idea of Taqqu's~\cite{TAQQU2011216}: use a 100-times-thicker SFHe layer, thus needing no muon cooling, so potentially providing two orders of magnitude higher intensity than the ``muCool" beam discussed above; it could be developed at Fermilab in parallel to the work in progress at PSI. This ``thick-film" approach could enable a $\stackrel{<}{_\sim}$\,$10$\% measurement of \gbar\ in a month of beam time at PSI~\cite{Antognini:2018nhb}, and potentially a 1\% or higher-precision measurement at a future Fermilab facility. Since only Mu atoms formed close to the upper SFHe surface will emerge upwards to form the desired beam, an electric field is maintained in the helium (via a pool of negative charge at the SFHe surface) to cause the stopping $\mu^+$ to separate from their ionization trails and drift to the upper surface before forming Mu. 
The $\sim$\,cm-wide beam results in some acceptance loss if cm-wide gratings are employed, thus larger gratings (if feasible) could be beneficial; alternatively, the SFHe deflector could have a curved surface so as to produce some focusing of the beam into the interferometer~\cite{Luppov:PhysRevLett.71.2405}.

Surface muon beams, available at J-PARC and MuSIC in Japan, ISIS in the U.K., TRIUMF in Canada, and PSI, are currently unavailable in the U.S. As the record holder for surface-muon beam intensity, PSI\,---\,with up to $\sim$\,$10^9$\,Hz surface-muon rate, and an upgrade to $10^{10}$ under discussion, to be produced using $\sim$\,$10^{12}$\,Hz of 590\,MeV protons on target\,---\,has been the natural venue for muonium-beam R\&D.
With potentially $\stackrel{>}{_\sim}$\,$10^{13}$\,Hz of protons on target, the coming PIP-II intensity upgrade~\cite{Shiltsev:2017} could make Fermilab the world leader for both fundamental muon experiments and the Muon Spin Rotation community~\cite{LEmuon-LoI};
 the novel muonium beams discussed above could be used as-is for MAGE~\cite{Antognini:2018nhb} and other muonium experiments, or ionized to serve
muon experiments~\cite{KANDA2014212}.

\vspace{-.1in}
\subsubsection{R\&D}\vspace{-.05in}

To enhance  beam design progress in the interim period before a new facility can be built, an R\&D platform would be extremely useful and, for some applications (e.g., SFHe Mu production), even crucial. This could be provided at  the Fermilab ``MuCool Test Area" (MTA),\footnote{A more ambitious scheme for a muon beam in the MTA is discussed in Ref.~\cite{Johnstone:IPAC2018-MOPML032}.} or (at lower intensity) using the Fermilab Test Beam Facility (FTBF). Other options may also be available.

\vspace{-.1in}
\subsubsection{Conclusion}\vspace{-.05in}

We propose to study the options for providing competitive muonium beams at Fermilab in the Mu2e and PIP-II ``eras." This study can inform  proposals for MAGE at Fermilab~\cite{MAGE-LoI} as well as other future experiments employing muonium, such as the precision determination of the  hyperfine and 1S--2S transition frequencies~\cite{gorringe2015precision,Mu-Mass,Mills:2014JPSPC...2a0401M}, the search for Mu--$\overline{\rm Mu}$ oscillation~\cite{Jungmann:2016}, etc.

The gravitational acceleration of antimatter, \gbar, has yet to be directly measured; an unexpected outcome of its measurement could change our understanding of gravity, the universe, and the possibility of a fifth force. Three avenues are apparent for such a measurement: antihydrogen, positronium, and muonium, the last requiring a precision atom interferometer and novel muonium beam under development. The interferometer and its few-picometer alignment and calibration systems appear feasible. With 100 nm grating pitch, measurements of gbar to 10\%, 1\%, or better can be envisioned, and are the goal of the MAGE collaboration. These could constitute the first gravitational measurements of leptonic matter, of 2nd-generation matter, and possibly, of antimatter. The coming PIP-II and Booster accelerator upgrades could make Fermilab the world’s best venue for such an experiment.

\subsection{Gravitational Effects on CP Violation} \label{section-kaon-system}


\subsubsection{Introduction}

Here, we consider a possible indirect measurement of antimatter gravity via a measurement of the dependence in the magnitude of CP violation as a function of gravitational field intensity.
To motivate the value of such an experiment, we note that gravity-generated CP violation could potentially help to explain ``missing'' antimatter in the universe (cosmic baryon asymmetry). Sakharov's conditions are satisfied in the Standard Model (SM)~\cite{gavela-1994a, gavela-1994b, PhysRevD.51.379}, while many non-SM theories imply a large CP violation and antigravity~\cite{
benoitlevy:2012,chardin-1993, villata:2011}.
In 1961, Good~\cite{good-1961} calculated that a repulsive gravitational interaction of antimatter should introduce a regeneration of kaons thus resulting in an anomalously large level of CP violation, at that time unknown. 
Chardin~\cite{chardin-1993} reformulated Good’s argument and showed that the gravitational field on the surface of the Earth is of the required order of magnitude to cause CP violation during the mixing time. Specifically, the mixing time of the $K^{0}$-$\bar{K}^{0}$ system, 
$\Delta\tau = 5.9 \times 10^{-10}$~s~$\simeq 6\tau_{K_S}$,
is long enough for the gravitational field of the Earth to attract the matter and repel the antimatter components of the $K$ meson to induce a separation, 
$\Delta\zeta = g(\Delta\tau)^2$, between them. When compared to the Compton wavelength of the kaon we obtain an adimensional measure of the phenomenon on Earth, 
$\chi = \Omega \times 0.88 \times 10^{-3}$ which is the same order of magnitude as epsilon. If we calculate $\chi$ given the gravitational strength on the Moon's surface, we expect the measured effect to be $\sim$97\% smaller than the effect measured on Earth's surface, assuming a linear dependence of the CP violation parameter, $\varepsilon$, with the gravitational acceleration (as in the case of repulsion between matter and antimatter~\cite{chardin-1993, good-1961}.
\begin{wrapfigure}{r}{0.41\textwidth}
  \begin{center}
  \vskip+40pt
  \includegraphics[width=0.40\textwidth]{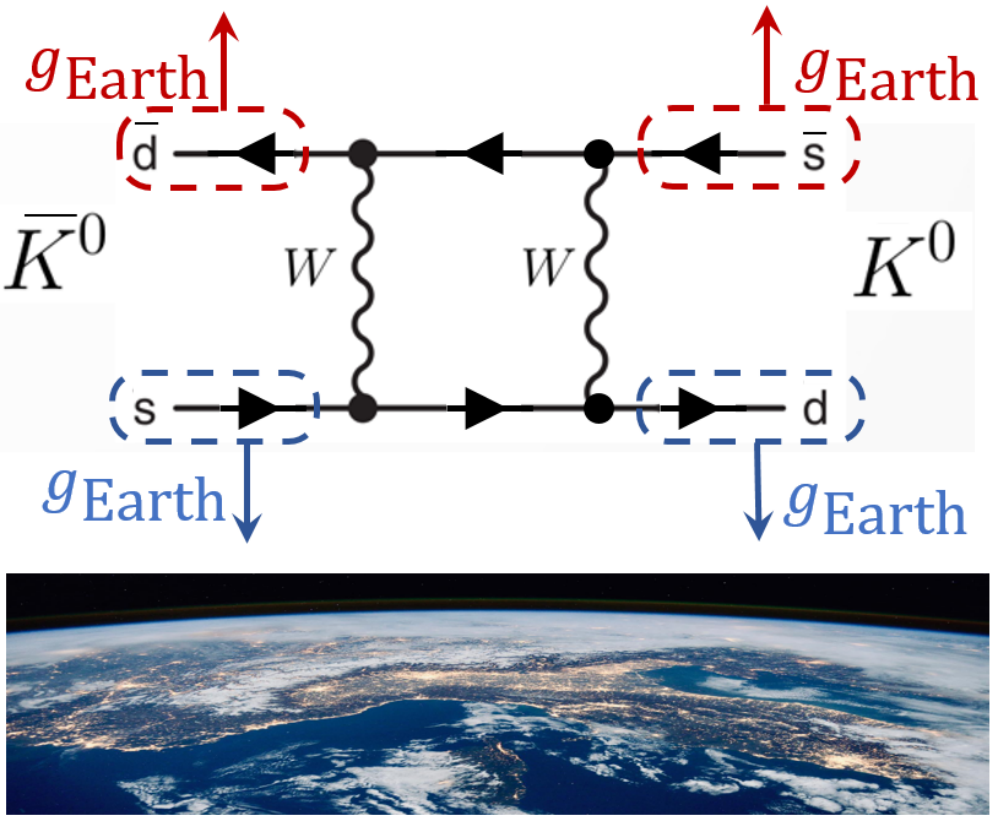}
  \end{center}
  \caption{\label{fig:NeutralKaon-FeynDiag}Within the neutral kaon system, the matter components could be attracted to Earth while the antimatter components are repelled.}\vskip-30pt
\end{wrapfigure}

\subsubsection{Experiment Concept} 
We propose to measure a dependence in the magnitude of CP violation as a function of gravitational field intensity. An experiment in Low Earth Orbit (LEO) would provide an environment with
$g_{LEO} = 0.9g_{Earth}$ while the surface of the Moon would provide an environment with
$g_{Moon} = 0.165g_{Earth}$.
In LEO or on the surface of the Moon where, due to the lower gravity, 
$R = \Gamma(K_L \to \pi^+\pi^-)/\Gamma(K_L \to \pi^+\pi^-\pi^0)$ is expected to be reduced by $\sim$20\% or $\sim$97\%, respectively. To produce the $K_L$ in either environment, one can leverage the flux of cosmic protons in place of the particle accelerators typically used in traditional experiments. A direct measurement of the flux of protons on the lunar surface has not yet been made, but the Cosmic Ray Telescope for the Effects of Radiation (CRaTER) aboard the Lunar Reconnaissance Orbiter~\cite{https://doi.org/10.1002/swe.20034, PhysRevD.93.082001}
measured the gamma albedo from the Moon surface due to the incoming cosmic proton flux and found it to be equal, within a 10\% uncertainty, to the proton flux measured by AMS-02~\cite{PhysRevLett.110.141102, consolandi2014primary}
and PAMELA~\cite{PAMELA:2017bna},
both in LEO. 
Piacentino et al.~\cite{Piacentino:2021xzt}, performed a Geant4 simulation with this spectrum of cosmic ray protons originating on a hemispherical surface with cosine-law biasing and incident upon a cylindrical target. The simulated apparatus consisted of a partially active cylindrical target with alternating layers of lunar regolith and scintillating material for a total depth of 18~cm. Simulations of an active target using using layers of PbWO$_4$, for a LEO experiment are described in their previous study, described in~\cite{piacentino-2016, piacentino-2017}).
They studied the amount of $K_L$ that would decay inside various sizes of downstream cylindrical tracking regions where the decay could potentially be reconstructed; for our initial estimate we used a reconstruction efficiency equal to 1 inside the fiducial volume. 
Table 1 shows the estimated the number of $K_L$ decays inside a 1~m radius 4~m deep cylindrical tracking volumes with an offset between the target and the tracking volume of 2~m to allow the $K_S$ to decay. Much of the remaining $K_S$ background contamination can be significantly reduced by selecting only $K_{S,L}$ that decay with low forward momentum (e.g., $p_z < 1$~GeV ) with minimal loss in the number of signal $K_L$ decays, as described in~\cite{piacentino-2016, piacentino-2017}. 
The additional background from misidentified $K_L \to \pi\mu\nu$ decays will be rejected with kinematic cuts during data analysis. Table 1 also lists the minimum amount of time it would take to collect a sufficient number of $K_L$ for 3$\sigma$ and 5$\sigma$ measurements of $R$, in each environment, with (and without) an assumed gravitational dependence on the CP violation parameter, $\varepsilon$. 

\subsubsection{Conclusion}
The environments in orbit around the Earth and on the surface of the Moon have numerous features (vacuum conditions, low gravity, and exposure to a relatively intense irradiation of cosmic protons covering a large spectrum of energy) that make them interesting not only for the study of astrophysical phenomena, but also for particle physics. We suggest an experiment sensitive to a possible difference between the amount of CP violation as measured on the surface of the Earth and in a lower gravity environment.
By placing a detector in either Low Earth Orbit or on the surface of the Moon, one could perform a direct measurement of the ratio of the number of $K_L$ decaying to two charged pions to those decaying to three pions in a low-gravity environment. It is estimated that it will take $\mathcal{O}$(days) to record sufficient $K_L$ decays for a 3$\sigma$ measurement of $R$, and $\mathcal{O}$(tens of days) for a 5$\sigma$ measurement. For the experiment on the Moon, if there is a dependence of $\varepsilon$ on $g$, within the first $\mathcal{O}$(tens of days) we would expect to measure only backgrounds, with a null signal measurement confirming the existence of a gravitational dependence. Any difference between the amount of CP violation in a low gravity environment with respect to the level CP violation on the surface of Earth could be an indication of a quantum gravitational effect.

The discovery of a gravitational dependence on the level of CP violation is sure to represent a significant milestone in our knowledge of particle physics. 
Its implication of the presence of a gravitational repulsion between matter and antimatter would constitute a systematic effect, not measurable in a laboratory on Earth, potentially capable of influencing the results of many high-energy experiments performed up to now. 
Such a discovery may motivate the subsequent development of a dedicated laboratory in space to repeat, under suitable gravitational conditions, a long series of experiments for which their Earth-surface based results may contain hidden gravitational contributions. 
The United States could be well-positioned to take a leading role in this endeavor by inaugurating a new and revolutionary line of space-based particle physics investigations. The experimentation should be carried out in a low-gravity environment, e.g., in Earth orbit, in lunar orbit, on the lunar surface, or elsewhere in our solar system~\cite{moon-village-2020}. While the International Space Station (ISS) has only nine years before it is slated to be decommissioned, its availability could be an important advantage. In fact, preliminary investigations and measurements could be carried out on the ISS to help inform the development of a detector for this proposed experiment. 



\section{Tests of general relativity and quantum effects related to gravity}

\subsection{Th-229 Nuclear Clock}
\label{Th}

\subsubsection{Introduction}
Currently, atomic clocks have a precision of a few parts in $10^{18}$ \cite{brewer:2019}. 
Time keeping this precise is generally done by measuring the frequency of an optical hypefine transition between two angular momentum 0 states.
Ultra-precise time-keeping has the potential to reveal new physics (ie: tests of the constancy of the fine structure constant, improved precision for tests of general relativity). A fluke of nature \cite{helmer:1994}, in Thorium-229, may lead to a potential development that could improve this precision by a factor of 100-1000 times. There is a nuclear transition that has an energy low enough ($\sim$7.8 eV) that could be directly excited by a laser at an approximate wavelength of $\sim$160 nm \cite{beck:2007}. By locking the laser frequency to the nuclear transition, one could create the world’s most precise nuclear clock by 2-3 orders of magnitude compared to the current state of the art. 

This low-lying nuclear level in 229Th has attracted the attention of scientists all over the world and has been the subject of much experimental and theoretical interest. Other research groups around the world have performed challenging experiments to study the properties of this isomeric state, including performing collinear laser spectroscopy on 229Th ions to study the hyperfine interaction, photon counting 229Th atoms guided to a target using a radiofrequency ion guide and buffer gas technique, and bombarding the 229Th atoms with intense x-ray beams from the Advanced Photon Source at Argonne National Laboratory. The 229mTh nuclear half-life has never been measured, and calculations are unreliable, ranging from microseconds to hours. Recently, the neutral-atom half-life has been inferred from the internal-conversion (electron signal) decay of 229mTh and found to be 7 $\mu$s \cite{seiferle:2017}. While this is a positive step forward, the critical knowledge of the energy to a precision needed for laser excitation and the half-life of the 229mTh nuclear state still remains.

\subsubsection{Experimental Concept}
Th-229m at a mere $7.6 \pm 0.5$~eV corresponds to a wavelength of approximately 160 nm and the transition has a spin difference of 1 h-bar, and the excited state is meta-stable with a half-life as long as hours. This makes 229Th the premier candidate for applying atomic spectroscopy techniques to a nuclear transition; ultraviolet-visible spectrometers could be used along with tabletop lasers and/or vacuum-ultraviolet (VUV) light sources to interrogate and to drive the transition between the two states of this nuclear doublet. 
The ability to apply the arsenal of precision optical spectroscopy techniques to the nuclear domain would be a breakthrough on par with the Nobel prize winning work of Mössbauer. Optical manipulation of the 229Th nucleus could lead to unprecedented studies of the interplay between atomic and nuclear systems, provide a new frequency/time standard, 
be used as a qubit for quantum computing with extremely long decoherence times, improve the search for time-variation of fundamental constants by as much as four orders of magnitude, 
and demonstrate for the first time coherent control of a nucleus.\\

In order to isolate the Thorium-229m isotope, an ion trap could be used to trap and confine Th-229m ions. By loading the ions into an ion trap with a high open solid angle, the ion trap can be readily observed for decay of the isomeric state. Lasers tuned to appropriate atomic transition wavelengths of the trapped ions could be used to non-destructively measure the trap population. Once a suitable population of ions has been trapped the isomer can be studied.

\subsubsection{Conclusion}
If the exact transition wavelength in Thorium-229m can be determined, a nuclear clock could be created utilizing the transition wavelength between the ground state and the isomer. This would potentially create a new international time standard, enable a host of general relativity experiments with unprecedented sensitivity, and enable an ultra-precise test of the constancy of the fine structure constant.

\subsection{Mechanical tests of the gravity-quantum interface}
\label{GQI}

\subsubsection{Introduction} Is gravity quantum? Apart from aspiring towards conducting experiments at the Planck energy scale, another way to address this question is to use low energy probes \cite{Carney_2019}, for example by attempting to gravitationally entangle two masses prepared in quantum states of their motion (see also a corresponding theory white paper submitted to Snowmass 2021  \cite{qgravtheorysnowmass}). If they are gravitationally entangled, then gravity must be quantum, if not, gravity must decohere their quantum state. 
Two classes of experiments can detect or falsify the presence of gravitational entanglement: interferometric tests that rely on preparing masses in a quantum superposition of their positions \cite{Feyn57,Bose:2017,Marletto:2017}, which would dramatically decohere when exposed to classical gravity; or non-interferometric tests that hope to precisely account for and measure the subtle effect of gravitational 
entanglement \cite{Diosi1989,penrose}.
The basic requirement in either case is the preparation and measurement of quantum states of motion of a solid-state mechanical oscillator. Atom interferometers have also been proposed as a way to infer the generation of gravitational entanglement \cite{Carney2021atom}.

\subsubsection{Experimental approaches}
Optomechanical systems have been identified as a promising route towards investigating the role of gravity in the entanglement of quantum systems \cite{carney2020mechanical,Carney_white_paper,carney2022newton, matsumura2020gravity, Bose:2017, Marletto:2017}. In this white paper we describe two examples of promising experimental methods, including interferometric and non-interferometric techniques. 

\textbf{Interferometric tests with levitated nano-particles.} From general relativity, mass generates curvature in spacetime and thus quantum mechanics should allow for quantum superpositions of different space-time curvature and for the gravitational field to mediate quantum entanglement between massive objects.  By developing new methods based on interferometry with levitated nanoparticles, despite the weakness of gravity, the phase evolution induced by the gravitational interaction of two levitated neutral test masses in adjacent matter-wave interferometers could detectably entangle them via graviton mediation even when they are placed far enough apart to keep other interactions at bay. Specific experimental proposals have been presented for using macroscopic superpositions of levitated nanoparticles to test whether the gravitational field can entangle the states of two masses \cite{Marletto:2017,Bose:2017},  
e.g. where embedded spins in the masses can be used as a witness to probe the entanglement \cite{Bose:2017, KimPRA2020}. Such experiments require an ultra-high-vacuum ultra-low-vibration cryogenic environment to minimize spurious environmental perturbations and technical noise.

The first specific experimental proposal for searching for a gravitational entanglement between two masses that are each in a superposition was based on nitrogen-vacancy centers (NVC) in diamond \cite{Bose:2017}. This is based on earlier proposals that an NVC in a spin superposition inside of a levitated nanodiamond in an inhomogeneous magnetic field could be used to create a macroscopic spatial superposition \cite{ScalaPRL2013, YinPRA2013, WanPRA2016}. To reach a large enough superposition distance it would probably be necessary to drop the nanodiamond as the trapping force tends to oppose the force creating the spatial superposition \cite{WanPRL2016}. Motional dynamic decoupling could be used to further increase the superposition distance and to remove many sources of decoherence \cite{PedernalesPRL2020}. This would also provide some NVC spin dynamic decoupling but much more would be needed which could be achieved by having the nanodiamond fall past magnetic teeth \cite{WoodPRA2022}. A Casimir screen could be put between the two nanodiamonds to reduce the unwanted Casimir interaction, making it easier for gravity to be the dominant interaction between the nanodiamonds \cite{vandeKampPRA2020}. Having the entire experiment housed in a freely falling platform, such as within a drop tower, could greatly reduce the relative acceleration noise \cite{TorosPRR2021}. Extensions of these proposals exist, such as aiming to close the loopholes in the Bell tests \cite{KentPRD2021}. 

Specifically, following the recent success in Stern-Gerlach (SG) interferometry with cold atoms on an atom chip \cite{Amit_2019, Margalit_2021}, an experimental roadmap has been outlined for an apparatus in which SG forces (i.e., magnetic gradients) applied to nanodiamonds holding a single embedded spin (in the form of a nitrogen-vacancy center), enable to put large masses in a spatial superposition \cite{Margalit_2021}. Recent feasibility studies have shown that this is doable \cite{Margalit_2021, Marshman_2021}, even if we take into account additional degrees of freedom such as phonons \cite{Henkel_2021} and rotations \cite{Japha_2022}. This opens the door for the numerous challenging theoretical proposals noted above.

A new class of more sensitive detectors could be built based on such a technology for detecting gravity, magnetic fields, electric fields, tilt and acceleration. More ambitiously, it has been proposed that a compact gravitational wave detector could be built in this way \cite{MarshmanNJP2020}. 

\begin{figure}[t!]
    \centering
    \includegraphics[width=\textwidth]{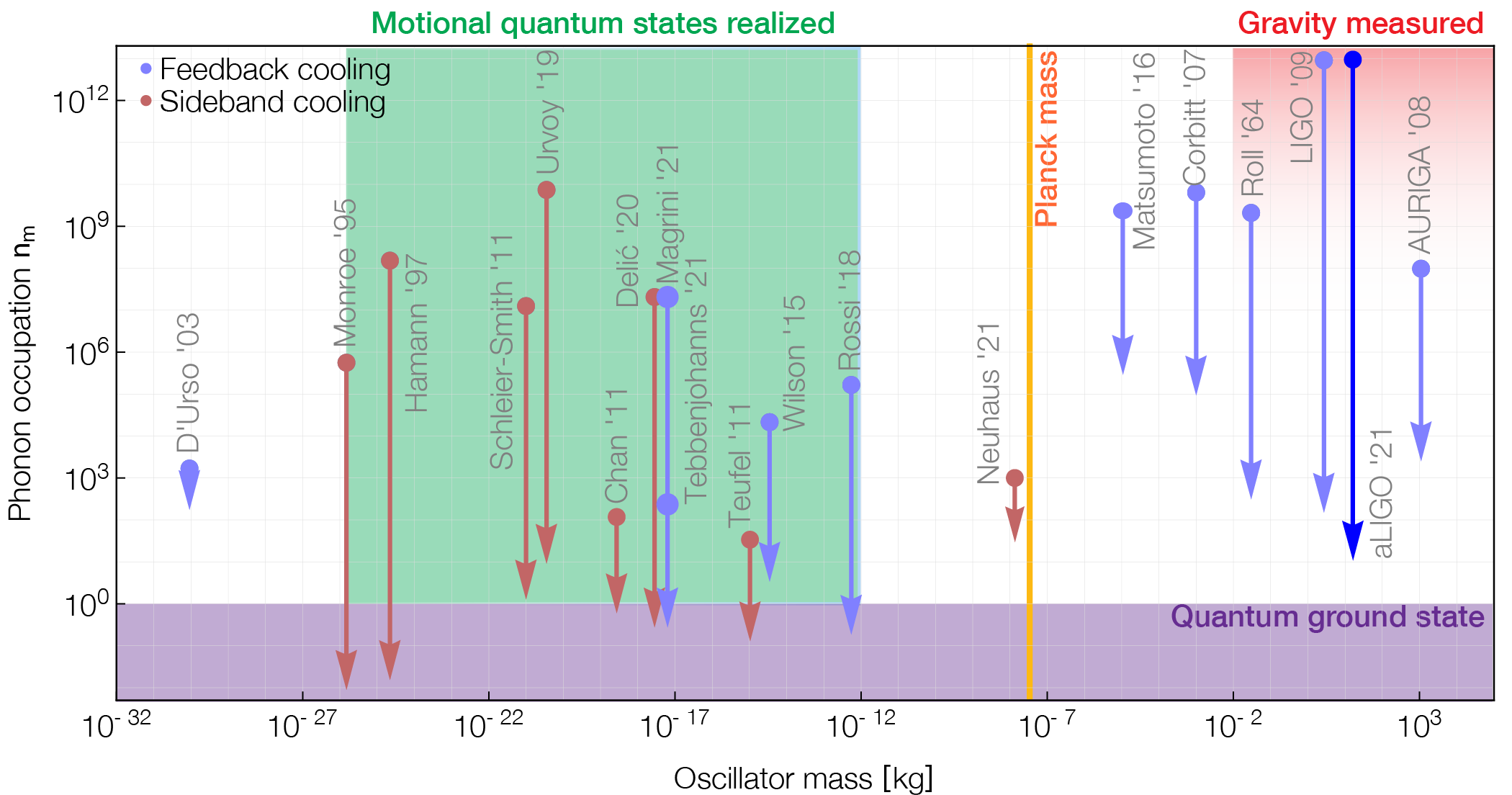}
    \caption{Contemporary survey of mass scales over which mechanical oscillators have been prepared in pure quantum states of motion. 
    These are all confined to the sub-nanogram scale. 
    Top right corner --- kilogram-scale masses in classical thermal 
    states of motion --- is
    the regime where classical gravitational interaction between
    masses has been observed.}
    \label{fig:cooling}
\end{figure}

\textbf{Non-interferometric tests with massive oscillators.}
In contrast to interferometric tests that require highly non-classical motional states of massive mechanical objects, non-interferometric tests aim to harness the remarkable sensitivity with which mechanical displacements can be measured to test the effect of gravity on
massive quantum systems. The fundamental challenge in a non-interferometric test of gravity's effect on
a massive quantum system is the preparation of nearly pure quantum states of motion of an oscillator that is sufficiently massive to appreciably gravitate with each other.
In the past decade, it has become possible to prepare quantum states of motion of
nano-/micro-scale solid-state mechanical oscillators \cite{Conn10,Chan11,Teuf11,PetReg16,Rossi18,MagAsp21,TebbNov21} --- but these systems are too light to gravitationally interact at distances small enough that extraneous (non-gravitational) near-field effects do not dominate the interaction. 
On the other hand, precise measurements of classical gravity have been performed with 
gram- and kilogram-scale masses \cite{Adelberger09} in classical states of motion. Fig.~\ref{fig:cooling} depicts the dichotomy between the current
state of affairs and the gap that needs to be bridged to enable a non-interferometric test of gravit's quantum nature.
Very recently, a kilogram-scale mechanical oscillator has been prepared close to its motional quantum ground state through 
measurement-based feedback control \cite{WhitSud21}, bridging the gap in mass across which pure quantum states of a massive object can be prepared.
The techniques demonstrated therein make it plausible to enter the regime where gravity can be sourced from an object prepared in a pure quantum state; further, a test mass, similarly prepared, can be used as a probe of gravitational entanglement or decoherence \cite{Diosi1989,penrose}. Gravitational decoherence can be witnessed
using quantum-noise-limited measurements that resolve the quantum fluctuations
of either system, whereas gravitational entanglement between the oscillators 
can be witnessed through joint measurements of their motion \cite{DattaQST2021}.  

\subsubsection{Conclusion}

Precision non-interferometric tests using mechanical oscillators prepared in quantum states are poised to enter the regime where
gravity can be sourced and sensed using quantum objects. This state of advance is largely due to the recent progress
in understanding the operating principles and limits of quantum-noise-limited displacement measurement and control of mechanical motion at
the quantum level. A new generation of table-top experiments are being planned to set stringent bounds on gravity's ability to mediate
entanglement. (Third generation gravitational-wave observatories such as Cosmic Explorer \cite{CE21}, and space-borne detectors such as 
LISA \cite{LISA18} will also be able to set stringent limits on gravitational decoherence.)
Levitated nano-particles that meet the criteria for an interferometric test of gravity's quantum character will come online over the next
decade. These experiments will eventually be limited by the fall-time available on earth. However, they are a necessary test-bed for 
eventual space-borne interferometric experiments \cite{MAQRO}. 
All these experiments share the need to understand and develop experimental techniques of broader impact such as low-environmental noise, 
mitigation of thermodynamic noises (for example via low-noise cryogenics, materials science, and engineering), and
shaping of quantum noises (for example, via quantum-enhanced metrology and control).

\subsection{Testing the effects of gravity on quantum spins} How intrinsic spin behaves in a spacetime that is warped by a massive rotating body is an experimentally open question. Levitated magnets have been identified as a system that allows one to go beyond the so-called energy-resolution limit (ERL) \cite{Vinante2021} and may have enough sensitivity to conduct experiments resembling Gravity Probe B, however, with quantum spin rather than mechanical angular momentum \cite{Fadeev2021}.  Freely floating ferromagnetic gyroscopes have also been identified as a route to search for new fundamental physics including exotic spin-dependent forces \cite{Fadeev_ferrogyro2021}.

\section{Searches for short-range corrections to gravity and other physics beyond the standard model \label{ShortRangeBSM}}

\subsection{Introduction} 
There is a vast  16 order of magnitude disparity between the apparent energy scale of quantum gravity, and that of the other Standard Model (electro-weak) forces.  However, as a number of recent theories have suggested, important clues related to this “hierarchy problem” can be obtained in low-energy experiments, by measuring how gravity behaves at sub-millimeter distances \cite{GiudiceDimopoulos,add}. But the gravitational force between massive objects becomes weak very rapidly as their size and separation distance decreases, thus making ultra-precise measurements a necessity at sub-millimeter length scales. 
In this white paper we elaborate on a selection of experimental methods, including torsion pendulua, slow neutrons, and levitated sensors which can be used to search for short range corrections to gravity and other fifth forces as well as other physics beyond the standard model. This list is not exhaustive. For example other promising spin-based methods for searching for novel short-range spin dependent interactions including atomic magnetometry and magnetic resonance have been discussed in detail in another Snowmass white paper on ``Quantum sensors for high precision measurements of
spin-dependent interactions'' \cite{Snowmass_Spins}.

\subsection{Experimental approaches} 
\subsubsection{Searches for exotic short-range gravity, equivalence-principle violation involving ordinary and dark matter, and novel spin-dependent interactions with torsion pendulums}
\label{pendulum}

Sensitive torsion balances are a powerful and proven method for studying exotic short-range gravity \cite{Kapner:2007,Lee2020}, equivalence-principle violation involving ordinary and dark \cite{wagnerEPV,Shaw2022} matter, and novel spin-dependent interactions \cite{Terrano:2015}. They remain one of the most promising paths forward for these studies as their sensitivity continues to increase and the understanding of background noise and systematic errors from patch charges and other surface forces improves.

Current tests are often limited by two factors:

1) environmental vibrations can “kick” the pendulum exciting its fundamental (twist) and spurious (swing, bounce and wobble) modes. This is particularly in short-range tests where patch charges couple to the spurious modes producing noise that dominates at small separations and limits the minimum attainable separation.

2) time-varying environmental gravity-gradients limit equivalence-principle tests.

Both of these technical limiting factors could be addressed by a development of a suitable underground facility that was open to outside users.

\subsubsection{Searches for Exotic Short-range Corrections to Gravity and Weakly Coupled Spin-Dependent Interactions using Slow Neutrons}
\label{exotic}

The special properties of slow neutrons enable unique types of precision measurement. The electrical neutrality of the neutron coupled with its small magnetic moment and very small electric polarizability make it insensitive to many of the electromagnetic backgrounds which can plague experiments that employ test masses made of atoms. The ability of slow neutrons to penetrate macroscopic amounts of matter and to interact in the medium with negligible decoherence allows the quantum amplitudes governing their motion to accumulate large phase shifts which can be sensed with interferometric measurements~\cite{Nico05b, Dubbers11, Pignol:2015}. These features of slow neutron interactions have been exploited in several searches for possible new weakly coupled interactions of various types, including chameleon dark energy fields, light $Z^{'}$ bosons, in-matter gravitational torsion and nonmetricity of spacetime, axion-like particles, and exotic parity-odd interactions~\cite{Leeb92, Bae07, Ser09, Ig09, Pie12, Yan13, Lehnert14, Jen14, Lemmel2015, Li2016, Lehnert2017, Haddock2018b, Cronenberg2018}. This strategy can succeed despite the uncertainties in our knowledge of the neutron-nucleus strong interaction. In the slow neutron regime with $kR\ll1$ where $k$ is the neutron wave vector and $R$ is the range of the neutron-nucleus strong interaction, neutron-nucleus scattering amplitudes are dominated by $s$-wave scattering lengths which are accurately measured experimentally. This makes coherent neutron interactions with matter sufficiently insensitive to the complicated details of the strong nucleon-nucleus interaction that one can cleanly interpret and analyze searches for small, exotic effects.

In this brief note we preset neutron searches for exotic gravity as an example. Many theories beyond the Standard Model postulate short-range modifications to gravity which produce deviations of Newton's gravitational potential from a strict $1/r$ dependence. Example speculations include the idea of compact extra dimensions of spacetime accessible only to the gravitational field~\cite{Arkani98,Arkani99, Adelberger03, Frank2004} and the idea that gravity might be modified on the length scale of 100 microns corresponding to the scale set by the dark energy density~\cite{Adelberger09}. Many extensions to the Standard Model of particle physics produce weakly coupled, long-range interactions~\cite{Jae10, Antoniadis11}. Certain candidates for dark matter in the sub-GeV mass range can induce Casimir-Polder-type interactions between nucleons~\cite{Fichet2017, Brax2018} with ranges from nuclear to atomic scales. 


\begin{figure}[t]
\centering
  \includegraphics[width=8cm]{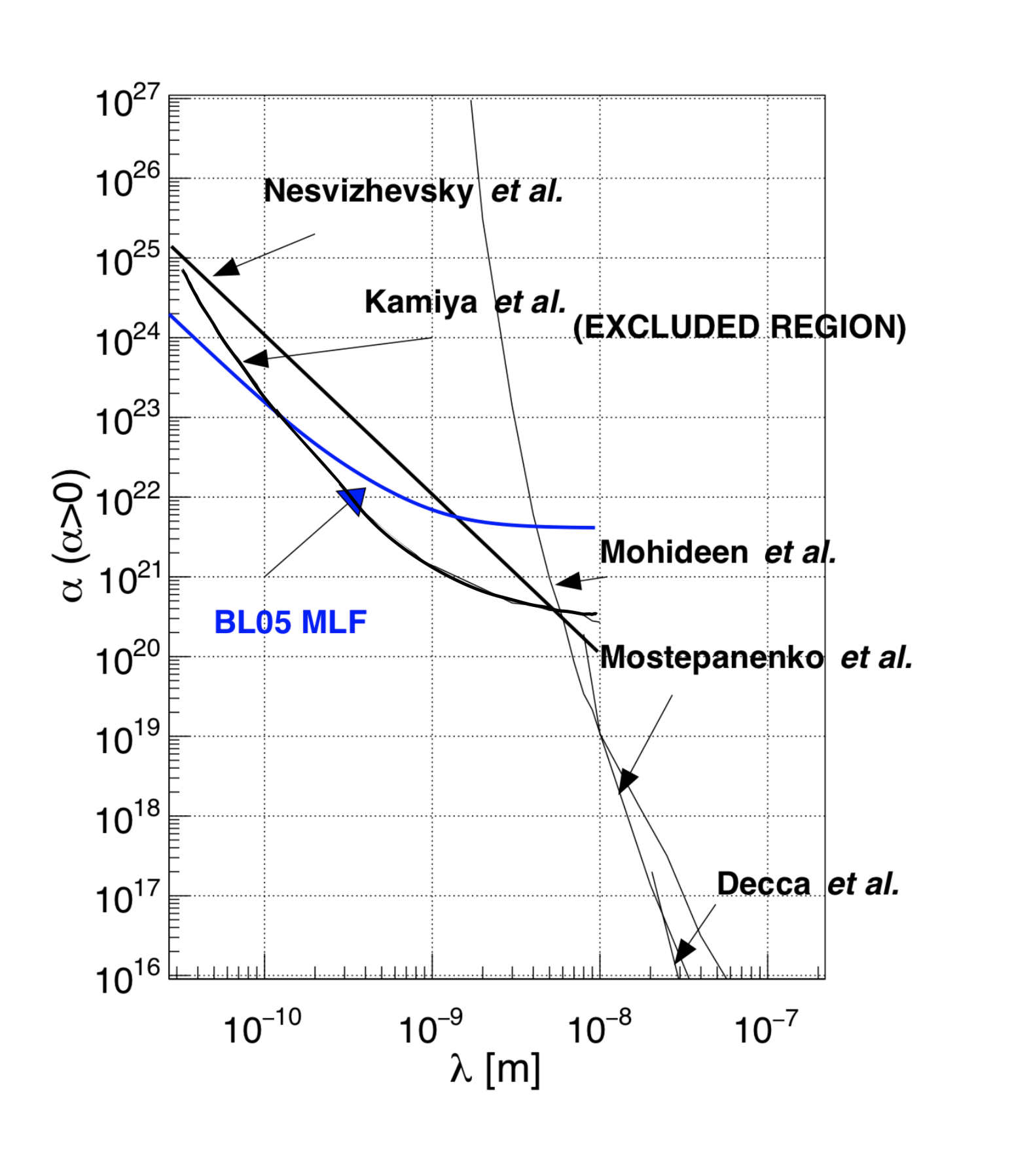}
\caption{Limits on the strength and range of short-range gravitational interactions of matter using neutrons and other probes in the $10^{-7}-10^{-11}$ meter range.}
\label{limitplot}
\end{figure}

It is common to analyze experiments searching for these modifications~\cite{Murata15} using a potential of the form $V^{\prime}(r)=-\frac{GMm}{r} [1+\alpha \exp{(-r/\lambda)}]$. The best present constraints on $\alpha$ for $\lambda$ between $10^{-8}$ and $10^{-13}$~m come from neutron scattering. Some constraints come from analysis of the neutron energy and $A$ dependence of neutron-nucleus scattering lengths~\cite{Nez08} measured to better than $0.1$\% accuracy for several nuclei. Other experiments have measured the angular distribution of neutrons scattered from noble gases to search for a deviation from that expected in this theoretically calculable system~\cite{Kamiya2015, Haddock2018a}. At shorter distances the best limits come from the measured energy dependence of neutron-nucleus cross sections in lead~\cite{Leeb92, Pokot06} and from very high-energy forward cross-section measurements at accelerator facilities~\cite{Kamyshkov08}.

The prospects for continued experimental progress are excellent. Ultracold neutrons are employed in gravity resonance spectroscopy~\cite{Jenke2011, Abele2008}, which creates coherent superpositions of bound states of neutrons formed in a potential from the Earth's gravity and a flat mirror. One can drive and resolve resonance transitions using acoustic transducers in a vibrational version of Ramsey spectroscopy. qBOUNCE has successfully conducted proof of principle measurements demonstrating vibrational Rabi spectroscopy~\cite{Abele2010}, and has sought several different types of exotic interactions~\cite{Ivanov2013, Jen14, Ivanov2016, Cronenberg2018, Klimchitskaya2019} through the influence of interactions sourced by the mirror material on the neutrons~\cite{Abele2003}. A new qBOUNCE apparatus which implements vibrational Ramsey spectroscopy has seen its first signal~\cite{Sedmik2019}. The GRANIT UCN spectrometer~\cite{Schmidt2009} at the ILL/Grenoble can conduct precision measurements on UCN gravitational bound states~\cite{Nesvizhevsky2002} with higher statistics when it is fed by a superfluid-helium-based UCN source~\cite{Piegsa2014}. With a bright very-cold neutron (VCN) source one could employ a Lloyd's mirror interferometer~\cite{Gudkov1993, Pokotolovskii2013a, Pokotolovskii2013b} to look for exotic interaction phase shifts from the mirror surface. Dynamical diffraction in perfect crystals can measure neutron scattering amplitudes at values of $q$ of about an inverse Angstrom and is sensitive to several types of exotic interactions \cite{greene2007neutron, voronin2009neutron}. The angular distribution of neutron scattering from noble gas atoms is sensitive to exotic Yukawa interactions through the $q$ dependence of the scattering form factor and measurements in progress at JPARC promise to better constrain exotic Yukawa interactions with ranges near the Angstrom 
scale. 


The great majority of neutron work in this area has made use of cold and ultracold neutrons, and there are many experimental opportunities for continued progress using neutrons in this energy range as most of these experiments are not yet limited solely by the statistical accuracy available in beams and sources at present neutron research facilities. However all of the arguments given above for the value of neutrons in this type of research also apply to neutrons of significantly higher energies. One of the new physical phenomena which appear in this energy regime are several sharp neutron-nucleus resonances, which are especially plentiful in heavy nuclei with their high level densities near the neutron separation energy. The much longer time (factors up to $10^{6}$) that a neutron spends in the nucleus in a resonance reaction compared to a potential scattering reaction provides an opportunity to greatly amplify the small effects of the exotic weakly-coupled interactions of interest. Many accelerator-based neutron sources developed for neutron scattering and materials science studies, nuclear spectroscopy needed for nuclear structure and reactions, astrophysics, nuclear fission, and applied nuclear data measurements make intense beams of epithermal neutrons. We anticipate that soon these beams will be used to conduct new types of searches for exotic interactions using neutrons.

\subsubsection{Optically levitated sensors for short-range gravitational tests}

Optically levitated dielectric objects in ultra-high vacuum exhibit an excellent decoupling from their environment, making them highly promising systems for precision sensing and quantum information science. In particular, the center of mass modes of optically-trapped silica nanospheres have exhibited high mechanical quality factors in excess of $10^7$ \cite{novotny2012} and zeptonewton ($10^{-21}$ N) force sensing capabilities \cite{ranjit2016}. Such devices make promising candidates for sensors of extremely feeble forces \cite{geraci2010}, accelerations \cite{andyhart2015, Moore2017,novotnydrop}, torques \cite{Li2016}, and rotations \cite{Li2018,Novotny2018,Moore2018}, testing the foundations of quantum mechanics \cite{oriol2011}, observing quantum behavior in the vibrational of modes of mechanical systems \cite{chang2009,coherentscattering, aspelmeyercavity}.

\begin{figure}[t]
    \centering
    \includegraphics[width=0.6\columnwidth]{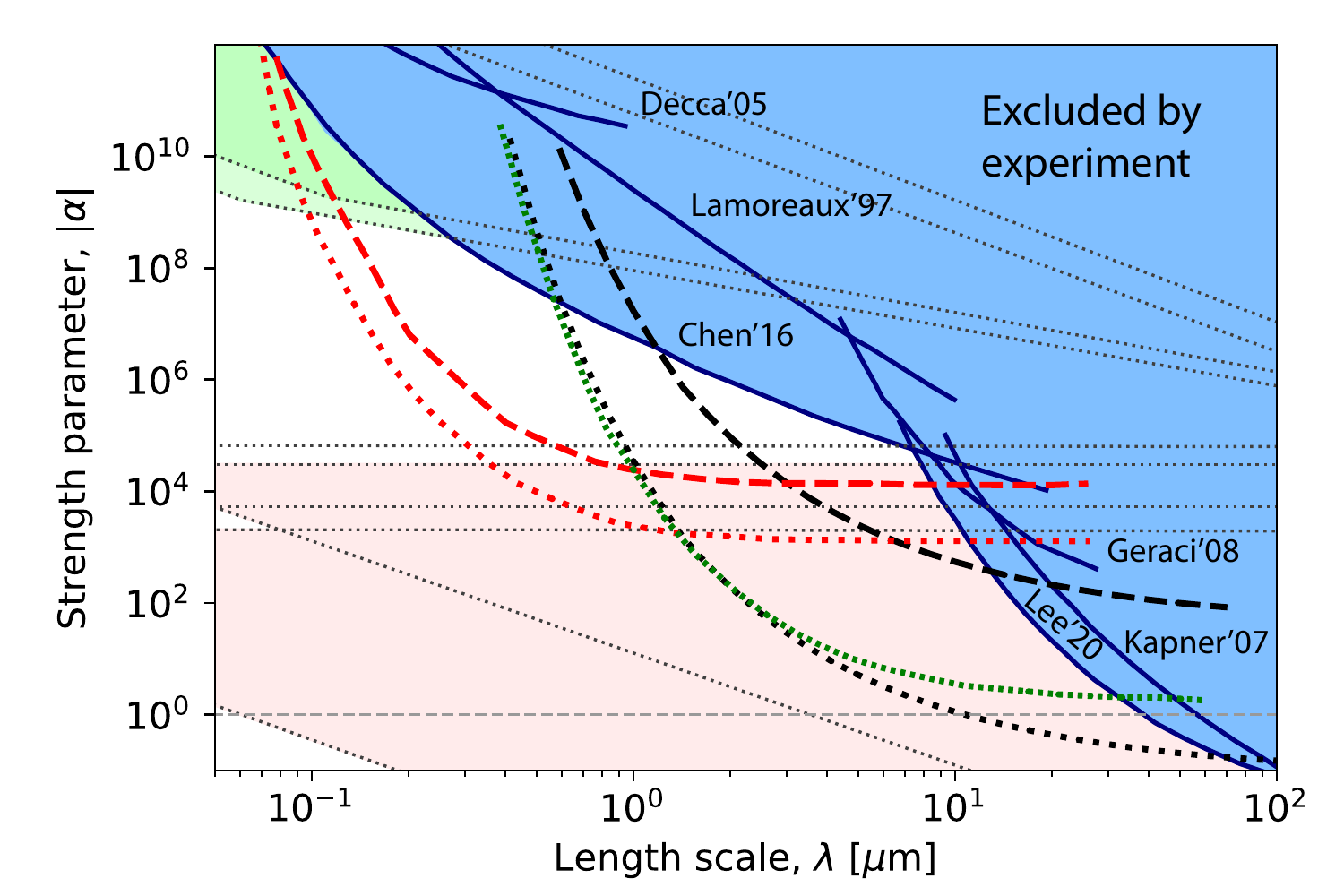}
    \caption{Adapted from Ref. \cite{GeraciMoore2020}. Background free sensitivity projections to Yukawa-type deviations of the form $V(r) = \frac{G_N m_1 m_2}{r} [1+\alpha \exp{(-r/\lambda})]$ from Newton's law for example optically levitated masses.  Existing limits are denoted by the blue region~\cite{Murata_2015,Deca:2016,Lamoreaux:1997,Decca,Kapner:2007,EotWash2020,Geraci:2008}, with allowed theory regions in a selection of models denoted in red and green~\cite{Geraci:2008}. The black dashed line shows the projected sensitivity for a 20~$\mu$m diameter sphere at the best currently demonstrated sensitivity for a sphere of this size~\cite{acceleration2020} for a $10^5$~s integration, assuming no backgrounds. The black dotted line shows the corresponding sensitivity at the Standard Quantum Limit.  The red dashed/dotted lines show the current/future sensitivity possible for a nanosphere with diameter of 300~nm~\cite{Geraci:2010}.  The green dotted line shows the projected sensitivity for a matter wave interferometer employing 13~nm diameter spheres~\cite{Geraci:matter_wave}. }
    \label{fig:gravity_sens}
\end{figure}

Trapped spheres can function as a test mass held using optical radiation pressure near the surface of an end mirror of an optical cavity. Non-Newtonian Gravity-like forces and Casimir forces can be tested by monitoring the motion of the sphere as a gravitational source mass is brought behind the cavity mirror. Other approaches involving an optical levitation trap are also being investigated \cite{Monteiro:2020wcb}. Several orders of magnitude of improvement is possible in the search for new gravity-like forces at the micron distance scale due to the sensitivity of the technique.  Fig. \ref{fig:gravity_sens} shows the potential reach along with theoretical predictions for new fifth forces that are Yukawa-type corrections to gravity at short distance scales using spheres of sizes 300 nm and 20 $\mu$m, currenently being investigated at Northwestern \cite{GeraciZepto}and Yale \cite{Monteiro:2020wcb}, respectively.

Advances in sensitivity made possible by pushing the sensitivity of these sensors into the quantum regime along with improved understanding and mitigation of systematic effects due to background electromagnetic interactions such as the Casimir effect and patch potentials will enable several orders of magnitude of improvement in the search for new physics beyond the Standard model.

\subsubsection{Other tests of fundamental physics with levitated particles} 

\textbf{Millicharged particles}

Levitated objects have a long history in testing the neutrality of matter and searching for fractionally charged particles. Ashkin first proposed the use of optically levitated spheres to perform a modern, ultra-sensitive version of the Millikan experiment in 1980~\cite{Ashkin1980}, and results of such an experiment were first reported in 2014~\cite{Moore:2014}. More recent results have provided the most sensitive search to data for particles with charges $\gtrsim 10^{-5}$~$e$ bound in terrestrial matter~\cite{Afek:2020}, as well as new approaches that can mitigate backgrounds coupling to electric dipoles in the particles~\cite{Priel:2021}. While searches to date are already sensitive to a charged relic dark matter component even if it makes up only a fraction of the overall relic density~\cite{Afek:2020}, future searches with larger masses can reach sensitivities to lower concentrations of such particles. Finally, similar techniques may allow tests of charge quantization and matter neutrality, with ultimate senstivity predicted to surpass the sensitivity of existing constraints~\cite{Afek:2020, Priel:2021}.

\textbf{Gravitational waves} The extreme force sensitivity made possible by optical levitation lends itself to the search for weak astrophysical signals, including feeble strain signals from Gravitational waves or impulses from passing Dark Matter.
One of the most interesting sources of Gravitational waves in the high-frequency regime arises from physics Beyond the Standard Model. The QCD axion is a well-motivated dark matter candidate that naturally solves the strong CP problem in strong interactions and explains the smallness of the neutron's electric dipole moment \cite{axion1,axion2,PTViolation,Moody:1984ba}. The Compton wavelength of the QCD axion with axion decay constant $f_a \sim 10^{16}$ GeV (at the Grand-Unified-Theory [GUT] energy scale) matches the size of stellar mass BHs and allows for the axion to bind with the BH “nucleus,” forming a gravitational atom in the sky. A cloud of axions grows exponentially around the BH, extracting energy and angular momentum from the BH \cite{stringaxiverse,stringaxiverse2}. Axions in this cloud produce gravitational radiation through annihilations of axions into gravitons. For annihilations, the frequency of the produced GWs is given by twice the mass of the axion: $f=145\ \mathrm{kHz}$, which lies in the optimal sensitivity range for optically leviated sensors when $f_a$ is around the GUT scale. The signal is coherent, monochromatic, long-lived, and thus completely different from all ordinary astrophysical sources. The fraction of the BH mass the axion cloud carries can be as high as ${10}^{-3}$ \cite{stringaxiverse2}, leading to strain signals detectable within the sensitivity band of optically levitated sensors \cite{GWprl}.  

\textbf{Dark Matter}
Dark matter can also be detected by observing the interaction of passing massive particles with the levitated nano-objects. For example, a recent search has been performed for composite dark matter particles scattering from an optically levitated nanogram mass, cooled to an effective temperature $\sim$200~$\mu$K~\cite{Monteiro:2020wcb}. Similar techniques may allow detection of sufficiently low momentum transfers that sub-MeV dark matter scattering coherently from 10~nm diameter spheres can be detected\cite{Afek:2021vjy}, or charged dark matter scattering from single trapped ions or electrons~\cite{Budker:2021quh,Carney:2021irt}. Large arrays of such trapped objects are possible, and can enable lower cross-sections to be reached~\cite{Afek:2021vjy,GeraciMoore2020}. Such detectors are intrinsically sensitive to the direction of the dark matter scatter, allowing an unambiguous determination of the astrophysical origin of a signal if detected~\cite{Monteiro:2020wcb,Afek:2021vjy,GeraciMoore2020}.
\subsection{Conclusion}

A variety of techniques including torsion pendulums, levitated optomechanical systems, slow neutrons, are ripe for research and development in order to extend the search for ``fifth-forces'' at short range, gravitational waves, dark matter, and other physics beyond the standard model. Research on improving sensitivity, ultimately harnessing quantum sensing techniques, and improving understanding and mitigation of backgrounds is needed to realize the full potential of these methods.

\section{Summary}

The coming decade provides numerous opportunities for significant advances in tests for fundamental physics by pushing the precision frontier in small- to mid-scale experiments. Tests of fundamental symmetries and gravity are particularly suited for a variety of precision techniques involving neutrons, anti-hydrogen, atomic clocks, matter wave interferometry, muon physics, penning traps, cavities, torsion pendulums and oscillators, optomechanical devices, and levitated particles. These platforms have been proven methods or have shown great promise and are ripe for investment in technological development. Going forward these methods are well positioned to extend the search for physics beyond the standard model by several orders of magnitude across unexplored parameter space.

\section{Corresponding Snowmass Letters of Interest}
This white paper has been assembled with input from the following LOIs submitted to Snowmass2021:
\begin{itemize}
    \item \href{https://www.snowmass21.org/docs/files/summaries/RF/SNOWMASS21-RF3_RF0-101.pdf}{Lorentz and CPT Tests with Low-Energy Precision Experiments~\cite{LOI1}}
    \item \href{https://www.snowmass21.org/docs/files/summaries/RF/SNOWMASS21-RF3_RF0_William_Snow-045.pdf}{NOPTREX: A Neutron OPtics Time Reversal EXperiment to search for Time Reversal Violation in Neutron-Nucleus Resonance Interactions~\cite{LOI2}}
    \item \href{https://www.snowmass21.org/docs/files/summaries/RF/SNOWMASS21-RF3_RF0-IF0_IF0_V_Sudhir-075.pdf}{Mechanical tests of the gravity-quantum interface~\cite{LOI3}}
    \item \href{https://www.snowmass21.org/docs/files/summaries/RF/SNOWMASS21-RF0-AF0-005.pdf}{Letter of Interest for a Muonium Gravity Experiment at Fermilab~\cite{LOI4}}
    \item \href{https://www.snowmass21.org/docs/files/summaries/RF/SNOWMASS21-RF3_RF0_Anthony_Palladino-056.pdf}{Letter of Interest for Snowmass 2021: Dedicated Experiment Exploring Gravitational Effects on CP Violation~\cite{LOI5}}
    \item \href{https://www.snowmass21.org/docs/files/summaries/RF/SNOWMASS21-RF3_RF0-CF2_CF0-IF1_IF0_Andrew_Geraci-076.pdf}{Optically levitated sensors for precision tests of fundamental physics\\ Snowmass LOI~\cite{LOI6}}
    \item \href{https://www.snowmass21.org/docs/files/summaries/RF/SNOWMASS21-RF3_RF0-128.pdf}{Searches for Exotic Short-range Gravity and Weakly Coupled Spin-Dependent Interactions using Slow Neutrons~\cite{LOI7}}
    \item \href{https://www.snowmass21.org/docs/files/summaries/RF/SNOWMASS21-RF3_RF0-CF7_CF0_Jason_Harke-046.pdf}{Th-229 Nuclear Clock~\cite{LOI8}}
\end{itemize}

\section{Endorsements}
In addition to the listed authors, the following people endorse and have expressed their support for this white paper:

\begin{itemize}
\item Michael E. Tobar,	Department of Physics, University of Western Australia, Crawley, 6009 WA, Australia
\item Yunhua Ding, W.M. Keck Science Department, Claremont McKenna, Pitzer, and Scripps Colleges, USA
\item Ronald Walsworth,	University of Maryland, USA
\item Hartmut Abele, TU Wien, Austria
\item Stefan Ulmer, Fundamental Symmetries Laboratory, RIKEN, 2-1 Hirosawa, Wako, 351-0198 Saitama, Japan
\item Klaus Kirch,	ETH Zurich and PSI, Switzerland
\item Sougato Bose,	University College London, UK
\item Anupam Mazumdar,	University of Groningen, Netherlands
\item Arnaldo J. Vargas, Laboratory of Theoretical Physics, Department of Physics, University of Puerto Rico, Río Piedras, Puerto Rico 00936
\item Jacob Dunningham,	University of Sussex, UK
\item Antonio Gioiosa,	University of Molise, Pesche, Italy INFN, Sezione di Roma Tor Vergata, Rome, Italy
\item Vincenzo Testa, Istituto Nazionale di Astrofisica - Osservatorio Astronomico di Roma, Italy
\item Fabrizio Marignetti, Università di Cassino e del Lazio Meridionale, Italy
\item Neil Russell,	Northern Michigan University, USA
\item Matthew Mewes, California Polytechnic State University, San Luis Obispo, CA, USA
\item Christian Ospelkaus, Leibniz Universität Hannover and Physikalisch-Technische Bundesanstalt, Germany
\item Jay D. Tasson, Carleton College
\item Christopher Haddock, Hedgefog Research
\item Philip Richerme, Indiana University
\item Robert D. Reasenberg,	Center for Astrophysics and Space Sciences (CASS), University of California San Diego, La Jolla, CA, USA \& Center for Astrophysics, Harvard and Smithsonian, Cambridge, MA, USA
\item Marianna S. Safronova, University of Delaware, Newark, DE, 19716, USA
\item F. Ignatov, BINP, Novosibirsk
\item Nicola Fratianni,	Università del Molise, Italy
\item Giuseppe Di Sciascio,	INFN - Roma Tor Vergata, Italy
\item R. N. Pilato,	Università di Pisa, Pisa, Italy; INFN, Sezione di Pisa, Pisa, Italy
\item Josh Long, University of Illinois at Urbana-Champaign, USA
\item Evan D. Hall,	LIGO Laboratory, Massachusetts Institute of Technology, USA
\item Eberhard Widmann,	Stefan Meyer Institute, Austrian Academy of Sciences, Austria
\item Chen-Yu Liu,	Indiana University, USA
\item Christian Schubert,	German Aerospace Center (DLR), Institute for Satellite Geodesy and Inertial Sensing, Callinstr. 36, 30167 Hannover, Germany
\item J.P. Miller, Boston University, Boston, MA, USA
\item Benjamin Heacock,	National Institute of Standards and Technology, USA
\item Livio Conti, Uninettuno University, Italy
\item Takeyasu Ito,	Los Alamos National Laboratory, Los Alamos, NM 87545, USA
\item Breese Quinn,	University of Mississippi, USA
\item Roberto Di Stefano, University of Cassino and Southern Lazio, USA
\item Alexander O. Sushkov,	Department of Physics, Boston University, Boston, Massachusetts 02215, USA
\item Tanja E. Mehlstäubler,	PTB, Braunschweig, Germany
\item Selim M. Shahriar,	Northwestern University, USA
\item Ivette Fuentes, University of Southampton, UK
\item Tim Kovachy, Northwestern University, USA
\item Daniel Carney, Lawrence Berkeley National Lab, USA
\item Nils A. Nilsson, Center for Quantum Spacetime, Sogang University
\item Tejinder P. Singh, Tata Institute Of Fundamental Research, Mumbai
\item Dennis Schlippert, Leibniz University Hannover, Institute of Quantum Optics, Welfengarten 1, 30167 Hannover, Germany
\item Quentin G. Bailey,	Embry-Riddle Aeronautical University, USA
\item Hendrik Ulbricht, University of Southampton, UK
\item Graziano Venanzoni, INFN Sezione di Pisa, Pisa, Italy
\end{itemize}

\bibliographystyle{Common/JHEP}
\bibliography{main}

\end{document}